# The Stability-Distillation Hypothesis for the Origin of Life

Cheng Bi

Email: bichengbj@ustc.edu

## Abstract

The field of abiogenesis harbors a logical impasse that has been tolerated for decades. Prevailing theories presuppose self-replication as the prerequisite for information accumulation, yet fail to explain the origin of replication itself. A self-replicating RNA molecule already embodies immense information content; its chance emergence from a random chemical soup approaches zero probability. For over forty years, this combinatorial explosion problem has never been directly resolved.

This paper advances a thesis fundamentally antithetical to the reigning paradigm: information precedes replication, rather than replication producing information.

We propose the mechanism of stability-distillation: under periodic environmental fluctuations—such as the wet-dry cycles of early Earth hot springs—molecules with stable conformations are preferentially retained while unstable ones are preferentially eliminated. The retained molecules serve as "seeds" for the next reaction round, their frequencies rising progressively across repeated cycles. Information thus arises spontaneously in a chemical system entirely devoid of replication capacity—the system transitions from a random state where all sequences are uniformly distributed, to an ordered state where stably folded sequences dominate collectively.

Computational simulations verify this process: within 1,000 wet-dry cycles, RNA sequences bearing stable stem-loop structures and closely resembling modern tRNA spontaneously emerge from random nucleotide monomers.

From this core mechanism, we derive seven propositions that are clearly distinguishable from existing theories and independently testable:

First, RNA was not "selected" as a genetic molecule but physically filtered as the most stable survivor. RNA became an information carrier not because it uniquely combines catalysis and information storage, but because base-pairing endows it with a distinctive "stability peak" distribution—the tiny minority of sequences capable of stem-loop formation are extraordinarily stable, while the vast majority of random coils undergo rapid hydrolysis. DNA, owing to its double-stranded structure, erases inter-sequence stability differences and thus cannot generate information via distillation; proteins, owing to their prohibitive combinatorial space, cannot converge independently at sufficient speed. Their respective roles were established progressively in subsequent evolution, not predetermined at the origin. Critically, RNA is not filtered in isolation: in realistic early Earth geochemical settings, heterogeneous molecules such as short peptides and lipids significantly reshape RNA stability distributions through non-covalent interactions. When specific short peptides bind to specific RNA sequences, both partners jointly gain stability advantages—the RNA provides a base-paired "stability scaffold," while the peptide gains indirect selective advantage through association. This RNA-peptide co-distillation implies that information and function accumulate simultaneously across multiple molecular species, rather than RNA information emerging linearly before protein function.

Second, the cell membrane is not a late-emerging "package" but a physical platform required from the very earliest stages of abiogenesis, defining a unified objective function that spans the entire history of life. The membrane's selective permeability—monomers can enter and exit, whereas macromolecules, once synthesized, are trapped—creates a fundamental selective pressure: any catalytic activity that polymerizes small monomers into macromolecular products unable to cross the membrane confers an osmotically driven growth advantage upon the host cell. This generalized macromolecule-production capacity constitutes the unified objective function of cellular evolution. RNA polymerase producing RNA, reverse transcriptase producing DNA, the ribosome producing proteins—these systems are functionally equivalent, all representing solutions that emerged sequentially under the same positive osmotic driving force. The driving force of evolution is "programmed" from the outset by the physicochemical properties of the membrane, requiring no genetic encoding or signal transduction. The cell is a parallel material-distribution computer, and the membrane itself is the hardware substrate of this computation.

Third, the ribosome is not a machine "designed" for translation but the product of "bricolage"—the piecing together of independent catalytic modules from metabolic networks. The ancestors of ribosomal protein components were free catalytic peptides that gained stability advantages by binding high-frequency stable RNAs within cross-catalytic networks. Before their integration into the ribosome, these peptides already possessed stable frequencies within metabolic networks and likely performed specific catalytic functions. Ribosome assembly is not teleological but a natural consequence of existing network components being joined together through cell fusion and material exchange.

Fourth, the primitive ribosome is functionally equivalent to a virus-like entity. Before the establishment and convergence of a DNA gene pool, the ribosome indiscriminately translates all accessible intracellular RNA, producing large quantities of useless proteins disconnected from the metabolic network. It hijacks host resources, induces osmotic imbalance and cell rupture, and propagates through the cell population via a "rupture-re-encapsulation" cycle—precisely matching the classical definition of a virus. The ribosome's transformation from "cell killer" to "cell factory" results not from altered intrinsic behavior but from two synergistic processes that reshaped its operational conditions: the convergence of the DNA pool eliminated RNA templates encoding useless proteins, while the translation system acquired selectivity toward RNA—only specific mRNAs are efficiently translated, the majority of non-coding RNAs being excluded. Purification at the template end and selectivity at the translation end reinforce each other, ultimately domesticating the ribosome from indiscriminate "viral-like burst" into controlled "factory production."

Fifth, cells and viruses do not stand in an antagonistic host–parasite relationship, but rather represent two distinct implementation paths under a unified objective function—the generalized capacity for macromolecular production. Cells adopt a "steady-state growth" strategy, achieving long-term accumulation through sustained macromolecular synthesis and controlled division. Viruses adopt a "pulsed outbreak" strategy, manufacturing macromolecules at an unsustainable rate over a short time window and spreading rapidly. The essential distinction between the two strategies lies in their optimization timescale: viruses maximize macromolecular production and transmission efficiency within a single infection cycle—a limit strategy targeting short-term fitness; cells pursue a sustainable balance between macromolecular accumulation and proliferative capacity across generational timescales—a steady-state strategy targeting long-term fitness. Both are jointly rooted in vesicle dynamics driven by positive osmotic pressure, representing optimal solutions of the same evolutionary algorithm operating on different timescales: when lipid supply is abundant, vesicles divide steadily (the ancestral cell); when macromolecular synthesis far outpaces lipid supply, vesicles rupture (the primordial viral cycle). The essence of a virus is not that of a "parasite," but rather a portable transmission format for highly efficient material distributions—an accelerator of the same evolutionary algorithm at the population level. Cells and viruses have coexisted from the very beginning of life's origin and have never truly been separate.

Sixth, LUCA was not a single cellular entity with a complete genome, but a dynamic gene pool undergoing active gene capture. The capture of ribosomal protein genes required their gradual acquisition through random mutation of cellular DNA, with dissemination and integration across the cell population via primitive viral vectors. The LUCA genome is precisely the fossil record of this process frozen mid-course. Recent independent reconstructions of the LUCA genome reveal a mature translation system yet a severe deficiency of ribosomal protein genes—we predicted this finding, rather than explaining it post hoc.

Seventh, sexual and asexual reproduction coexisted from the earliest stages of abiogenesis. Viral transmission is functionally equivalent to sexual reproduction—both represent horizontal recombination of material-distribution recipes already validated within different cells, and both share ancient molecular mechanisms. Asexual reproduction ensures the stability of vertical inheritance; sexual reproduction and its viral equivalent enable horizontal exploration and optimization. This binary architecture is not a late invention following the emergence of eukaryotes but a necessary division of labor from life's inception.

The seven propositions above form a logically self-consistent and complete framework: each step's conclusion serves as the premise for the next; each step's driving force derives from the same core principle—selective enrichment via stability differentials—unfolding naturally under different physical and chemical contexts. This framework does not negate existing hypotheses; rather, it provides a unified physicochemical foundation for the rational kernels each has captured.

Here are the testable predictions that are distinct from existing theories. Three core predictions jointly test the central deduction that "functional protein components can arise through co-distillation before the emergence of translation systems":

(1) In experiments simulating wet-dry cycles, random RNA libraries and random short peptide libraries should undergo bidirectional sequence convergence—RNA sequences should converge toward those that stably bind high-frequency peptides, while peptide sequences should converge toward those that stably bind high-frequency RNA. Existing experiments have shown that cationic short peptides can form mutually stabilizing complexes with RNA; this prediction further requires sequence-specific bidirectional convergence.

(2) Short peptide fragments from conserved domains of ribosomal proteins, when co-incubated with stable RNAs resembling rRNA precursors, should form more stable complexes than random peptides. In the RNA stability distillation simulations of this study, sequences with over 50% similarity to modern rRNA spontaneously and repeatedly emerged from random monomers—strongly suggesting that peptide sequences optimally adapted to such rRNA-like structures can be directly enriched through RNA-peptide co-distillation. The conserved core sequences and structural modules of ribosomal proteins very likely pre-existed through physicochemical co-stabilization mechanisms before the emergence of genetically encoded translation.

Some of these proteins, or the complexes they form with RNA, may have exerted independent catalytic functions in early cellular metabolic networks.

(3) Other proteins that must have existed before the translation system was fully established—including reverse transcriptase, RNA polymerase, and aminoacyl-tRNA synthetase—should contain short peptide fragments from their conserved domains that form sequence-specific stable complexes with their corresponding stable RNAs (tRNA or rRNA precursors). All three classes of proteins are traced by phylogenetic analysis to the earliest stages of life, all use stable RNAs as their natural substrates or interaction partners, and all are core components of the genetic information processing system. In conventional frameworks, this combination of features can only be described as an "evolutionary coincidence"; in the framework of this paper, it is a necessary consequence of stability distillation.

Several predictions have already received preliminary support from recent independent studies—including the severe incompleteness of ribosomal protein genes in LUCA, the abundance of ancient viral sequences in modern genomes, and the shared ancient molecular mechanisms between sexual reproduction and viral entry. The above co-distillation predictions remain to be experimentally tested.

If this framework is correct, the origin of life is no longer a fortuitous miracle dependent on ultra-low-probability events, but the inevitable emergence of a complex chemical system under specific boundary conditions. The decades-old quest to "find the first replicator" would be recast as having asked the wrong question from the start.

# Chapter 1: Introduction—Reframing the Problem of Life's Origin

## 1.1 The Core Puzzle: Where Does Information Come From?

What fundamentally distinguishes the living from the non-living? At the molecular level, the most salient feature of living systems is not the peculiarity of their chemical constituents—carbon, hydrogen, oxygen, nitrogen, and phosphorus are equally abundant in abiotic environments—but the non-randomness of their molecular composition.

In a living cell, thousands of macromolecular species coexist in precise proportions and specific sequences. This highly ordered molecular distribution stands in stark contrast to the uniform, disordered mixtures produced by any random chemical reaction. From an information-theoretic perspective, the fundamental difference between living and non-living systems lies not in the absolute entropy of the global molecular distribution but in its pronounced non-randomness: the frequencies of specific functional sequences far exceed the expectation from random chemistry. In a completely random chemical system, all sequences appear at comparable frequencies and no information exists regarding "which sequences matter." Living systems, by contrast, have evolutionarily filtered a specific set of functional molecules (e.g., rRNA, tRNA), maintaining them at persistently high copy numbers despite a background of abundant low-frequency random sequences.

The origin-of-life problem can thus be reframed as: how does a completely random chemical system, in the absence of life, spontaneously generate this statistical deviation of functional sequence frequencies from random expectation—i.e., how does it generate information?

The information-theoretic approach to the origin of life has a long scholarly tradition (Yockey, 2005; Walker & Davies, 2013). The difficulty of this problem stems from an apparently insurmountable physical barrier: the Second Law of Thermodynamics dictates that the entropy of an isolated system always tends to increase, while selecting specific functional sequences from a near-infinite combinatorial space in a noisy chemical environment constitutes a formidable combinatorial explosion.

## 1.2 The Predicament of Existing Hypotheses: The Paradox of Self-Replication as Starting Point

The two most influential theoretical frameworks for the origin of life—the RNA World hypothesis and the Metabolism-First hypothesis—each capture important facets of life's essence, yet neither truly answers the first-mover question of where information comes from.

The RNA World hypothesis posits that life originated from RNA molecules capable of both storing genetic information and catalyzing chemical reactions (Gilbert, 1986; Joyce, 2002). Its core predicament is this: a self-replicating RNA molecule is itself an information-rich, complex sequence whose chance emergence from a random chemical soup approaches zero probability. More critically, the hypothesis cannot explain how the capacity for replication itself arises de novo—without pre-existing replication, no specific sequence can be "remembered" or "amplified."

The Metabolism-First hypothesis proposes that life originated from a self-organizing network of chemical reactions (Wächtershäuser, 1988; Morowitz, 1992), correctly emphasizing the importance of compartmentalization and continuous energy flux. However, while it describes how metabolic networks self-sustain, it does not clarify how information is generated and accumulated within such networks—how the network transitions from "randomly producing diverse molecules" to "selectively enriching specific functional molecules."

Despite their different starting points, these two hypotheses share a deep presupposition: life must originate from some molecular entity capable of self-replication or self-catalysis. This presupposition constitutes a severe logical obstacle.

### 1.2.1 Self-Replication as the Paradox of Maximized Entropy Reduction

From an information-theoretic perspective, self-replication can be understood as the maximally efficient means of reducing information entropy. When a molecule can precisely replicate itself, its sequence frequency rises dramatically in the system, and the system's information entropy plummets. Yet this very perspective reveals the profound paradox of taking self-replication as the starting point: self-replication represents the terminal state of information entropy reduction, not the initial state. Setting a mechanism requiring the highest degree of information accumulation as the starting point of information accumulation constitutes an intractable circular argument.

### 1.2.2 The Physical Impossibility of an "Omnipotent" Macromolecule under Early Earth Conditions

If life originated directly from a self-replicating macromolecule, that macromolecule would need to simultaneously satisfy stringent conditions: (1) ultra-high stability under extreme conditions—high temperature, intense radiation, and frequent wet-dry cycling; (2) broad-spectrum catalytic capacity to coordinate substrate recognition, directed polymerization, and correct product folding; (3) sustained progression through multi-step directed synthesis—before the first complex self-replicating macromolecule could appear, a series of chemically irreversible assembly steps would have to occur, each error requiring a complete restart, yielding near-zero probability in a random environment; (4) continuous operation heavily dependent on ample, stable supplies of specific substrates from the surrounding milieu—pushing the question of information origin yet one step further upstream.

The individual probabilities of these four conditions are each extremely low; the probability of their simultaneous satisfaction is the product of these low probabilities.

Both hypotheses describe pieces of the puzzle but both miss the most critical piece: the transition mechanism from randomness to order. More fundamentally, both set a mechanism requiring the highest degree of information accumulation (self-replication) as the starting point of information accumulation.

## 1.3 Core Framework: Stability-Distillation and the Parallel Material-Distribution Computer

This paper proposes an entirely new theoretical framework whose central claim is: before natural selection could operate, a more primitive and fundamental sorting principle—stability-distillation—drove chemical systems from randomness toward order. The cellular structure then provided a computational platform for material distribution.

### 1.3.1 Stability-Distillation: A Replication-Independent Mechanism of Information Generation

Stability-distillation, the core concept introduced in this paper (Bi, 2024), refers to the physicochemical mechanism by which molecular stability differences are exploited under periodic environmental fluctuations to abiotically generate and accumulate information.

When a system containing diverse random macromolecules undergoes periodic environmental changes (such as wet-dry cycles), unstable macromolecules are preferentially degraded while stable ones are preferentially retained. In the next cycle, the retained stable molecules serve as fresh "feedstock" for subsequent synthesis and polymerization. With increasing cycles, the frequencies of stable molecules continuously rise while those of unstable molecules continuously fall—the system's molecular frequency distribution shifts from uniform to concentrated, and information entropy correspondingly decreases.

Stability-distillation bears a formal resemblance to Eigen's quasi-species theory (Eigen, 1971), but a fundamental difference exists: in quasi-species theory, "fitness" is replication rate, and its operation presupposes a population of self-replicating molecules; in stability-distillation, "fitness" is physicochemical stability, and the mechanism can operate spontaneously in chemical systems entirely devoid of replication capacity. This distinction is decisive: stability-distillation represents a gradual, cumulative process of information accumulation, where the system departs from a fully random initial state, with each step producing only minute frequency shifts that nevertheless accumulate into significant information over sufficiently many cycles.

### 1.3.2 The Cellular Structure as a Parallel Material-Distribution Computer

Stability-distillation alone is insufficient to explain the origin of life, because it can only filter molecules by their individual physicochemical stability and cannot select for the functional cooperativity of molecular networks. The transition to life requires a computational platform capable of actively exploring and optimizing combinatorial molecular functions.

That platform is the cellular structure. This paper argues that the primitive cell is an active, parallel material-distribution computer. Its computational logic comprises:
(1) Parallelism—a primordial hot pool simultaneously hosts vast numbers of cells, each an independent "laboratory" encapsulating distinct molecular combinations;
(2) Statistical inheritance—daughter cells need not perfectly inherit every molecule from the parent; inheriting a "representative sample" suffices to carry forward the functional advantages discovered by the preceding generation; (3) Population-level selection—cells that happen to encapsulate more efficient molecular combinations exhibit higher catalytic efficiency in their internal metabolic networks, are more prone to divide or rupture under thermodynamic drive, and thereby release their contents back into the environment, raising the frequency of these successful molecular combinations in the population; (4) Fusion-driven recombination—frequent fusion events between cells merge distinct molecular networks, potentially yielding novel combinations more efficient than either parent. This concept builds on established experimental and theoretical work on primitive cell population dynamics (Zhu & Szostak, 2009; Chen et al., 2004).

This "division–fusion–selection" cycle implements, at the population level, a distributed, physically driven parallel search algorithm capable of efficiently locating optimal solutions within vast combinatorial molecular spaces.

### 1.3.3 A Core Prediction: The Driving Force of Information Accumulation Lies in the Physicochemical Properties of the Membrane

An important corollary of this paper is that the evolutionary driving force for intracellular material distribution is ultimately rooted in the capacity to manufacture macromolecules that cannot freely cross the cell membrane. Because of the membrane's selective permeability—monomers have a significantly higher probability of crossing, but macromolecules, once synthesized, are trapped inside—intracellular macromolecular synthesis continually reduces internal monomer concentration, generating an osmotic gradient that drives continuous monomer influx from the environment.

This growth advantage is driven entirely by the physicochemical properties of the membrane, requiring no genetic encoding or complex signal transduction. The cell membrane itself is the hardware substrate of this evolutionary algorithm.

## 1.4 The Nature of Information: Persistence Duration and Statistical Significance

The theoretical framework of this paper rests on a core insight: the source of all information can be traced to a single fundamental physical fact—different molecules possess different stabilities and consequently different persistence durations. This concept physically corresponds to the half-life of radioactive elements. In a chemical system, the average survival time of an RNA molecule with a stable tertiary structure may be tens of times longer than that of a random-coil RNA.

It is precisely this difference in persistence durations that provides the physical basis for information. Information is defined here as the statistical deviation of functional sequence frequencies from random expectation—the greater the deviation, the higher the information content. This definition aligns with the reality of living systems: in modern cells, rRNA and tRNA account for 80–90% of total RNA mass, but the enormous diversity of mRNA species means the global Shannon entropy does not approach zero. The hallmark of life is not the elimination of all diversity, but the stable maintenance of high frequencies of specific functional molecules against a noisy background.

## 1.5 From Physics to Biology: A Graduated Evolutionary Landscape

Building on the two core concepts of stability-distillation and the parallel material-distribution computer, we progressively derive a complete evolutionary landscape from simple chemical systems to complex living systems:

(1) Environmental phase (RNA stability-distillation): In the wet-dry cycles of early Earth hot springs, RNAs with stable conformations are selectively retained and enriched through abiotic distillation, becoming carriers of primitive information.

(2) Cellular phase (establishment of metabolic networks): When lipid membranes self-assemble into primitive cells, compartmentalization provides molecular systems with a "privatized" concentrating space and information boundary. The coordinated operation of catalysts and degradative enzymes upgrades passive distillation to active filtering.

(3) Ribosomal phase (emergence of the translation system): On the foundation of complex metabolic networks, the first primitive ribosome with generalized translation capacity is assembled, marking the leap from "non-genetically encoded metabolic networks" to "programmable protein synthesis."

(4) Genetic encoding phase (information consolidation and revolution): DNA, via reverse transcription, "snapshots" RNA information into stable genetic sequences. The establishment of a genetic encoding system upgrades cellular material-distribution computation from "statistical recombination at the level of molecular frequencies" to "deterministic propagation at the level of gene sequences." Throughout this process, cells and primitive viral vectors undergo co-origin and co-evolution, ultimately forming the core architecture of modern genetic systems.

## 1.6 Structure of This Paper

The subsequent chapters unfold in accordance with the graduated landscape described above. Chapter 2 formally defines the principle of stability-distillation and demonstrates its operation in RNA systems via computational simulation. Chapter 3 argues for the inevitability of RNA as the primordial information carrier. Chapter 4 elaborates how cellular compartmentalization upgrades distillation into system-level computation. Chapter 5 describes how metabolic networks achieve non-genetically encoded information growth prior to the emergence of ribosomes. Chapter 6 sets forth how cross-catalytic networks drive the spontaneous assembly of the ribosome. Chapter 7 derives the establishment of genetic encoding systems and the co-origin of cells and viruses. Chapter 8 unifies the above mechanisms within a broader theoretical landscape, exploring the extension of stability-distillation to the transition from unicellular to multicellular life, its deep structural homologies with neural systems and artificial intelligence, and provides testable predictions and future research directions.

The logical arc of this paper can be summarized as a single chain of necessity: differential stability → spontaneous emergence of information → physical selection of RNA → compartmentalization as computational platform → non-genetic information accumulation in metabolic networks → ribosomal assembly from cross-catalytic modules → co-origin and coexistence of cells and viruses. Each link in this chain constitutes the premise for the next, and each transition is driven by the same underlying principle—selective enrichment via stability differences—operating under progressively more complex boundary conditions.

The aim of this paper is to demonstrate that, under plausible early Earth physicochemical conditions, the entire transition from random chemistry to genetic systems can be derived through a unified, logically necessary mechanism, without recourse to any ultra-low-probability chance events. If this argument holds, then the origin of life is no longer an inscrutable "fortuitous miracle" but the "inevitable emergence" of a complex chemical system under specific boundary conditions.

# Chapter 2: Stability-Distillation—The Physical Principle of Information Generation

## 2.1 Formal Definition of Stability-Distillation

Consider a chemical system containing diverse random macromolecules. Due to the statistical nature of random polymerization, any specific sequence constitutes an exceedingly low fraction of the total, and the system's information entropy approaches its maximum.

Now subject this system to a periodic environmental fluctuation. Each cycle consists of two alternating phases:

Synthesis phase (dry environment): The system is at high concentration and high temperature, promoting polymerization and ligation among monomers and macromolecules. Existing molecules in the system serve as "feedstock" for new polymerization, generating longer or more complex sequence combinations.

Degradation phase (wet environment): The system is at low concentration and low temperature; hydrolysis accelerates. Macromolecules with insufficient structural stability are preferentially degraded—their chemical bonds exposed to water attack, with sequences lacking protective folding being the first casualties.

Molecules of different sequences face starkly different fates: those with stable secondary or tertiary conformations—such as RNA stem-loop structures stabilized by base-pairing—can resist hydrolysis to some degree (Nowakowski & Tinoco, 1997; Serra et al., 1993), while random coils lacking stable conformations are rapidly degraded into monomers or short fragments.

When the system enters the synthesis phase of the next cycle, the surviving stable molecules become the "seeds" or "feedstock" for the next round of polymerization. Because their relative frequencies in the system are already slightly elevated above those of other sequences, the frequencies of their subsequent ligation and polymerization products are correspondingly higher. As these synthesis–degradation cycles repeat, the frequencies of stable molecules continuously rise while those of unstable molecules continuously fall—the system's molecular frequency distribution shifts from uniform to concentrated.

Definition of stability-distillation: A non-equilibrium process in which, under periodic environmental fluctuations, the system exploits molecular stability differences to repeatedly eliminate unstable components while repeatedly retaining stable components as "feedstock" for subsequent reactions, resulting in a progressively narrowing molecular frequency distribution and persistently increasing non-randomness.

## 2.2 Mathematical Description

### 2.2.1 Definition of Information Entropy

Consider a chemical system containing N distinct macromolecular sequences, with the frequency of the $i$-th sequence being $p_i$ (satisfying $\Sigma p_i = 1$). The system's information entropy H is given by the Shannon entropy (Yockey, 2005):

$$H = -\sum_{i=1}^{N} p_i \log_2 p_i$$

When all sequences have equal frequencies, the entropy attains its maximum $H_max = log_2 N$. When the frequency of one sequence approaches 1 and those of all others approach 0, the entropy approaches 0. From an information-theoretic perspective, the process of abiogenesis is one in which the system's molecular frequency distribution transitions from a near-uniform random state toward a state where the frequencies of certain specific sequences significantly deviate from random expectation. It should be noted that, during this process, the decline in global Shannon entropy is dampened by the continuous generation of low-abundance random background sequences—consistent with the fact that in modern cells, although rRNA and tRNA account for 80–90% of total RNA mass, the global Shannon entropy does not approach zero.

Accordingly, in subsequent discussions, we measure the information-accumulation effect of distillation with two complementary metrics: (1) the trend in global Shannon entropy, as a reference for the system's overall degree of order; (2) the degree to which dominant sequences deviate in frequency from random expectation, as the primary measure of information accumulation. In the simulation results, we additionally introduce "module convergence"—the gap between the actual frequency distribution of functional sequence fragments and random expectation—to more precisely quantify the information output of distillation.

### 2.2.2 The Evolutionary Equation of Stability-Distillation

We can describe the effect of stability-distillation on the molecular frequency distribution using a simplified evolutionary equation.

Let there be molecules $X_i$ (the $i$ -th sequence) in the system, with frequency $p_i(t)$ after the $t$-th cycle, and stability $s_i$ ($0 < s_i < 1$, with higher values indicating greater stability). In each cycle, the survival probability of molecule $X_i$ in the degradation phase is proportional to $s_i$; in the synthesis phase, surviving molecules serve as feedstock for polymerization and ligation, with the frequencies of newly generated polymerization products depending on the frequencies of their precursors (i.e., the surviving feedstock molecules).

The simplified kinetic equation is:

$$p_i(t+1) = \frac{s_i \cdot p_i(t) + \alpha \cdot g_i(\boldsymbol{p}(t))}{\sum_j [s_j \cdot p_j(t) + \alpha \cdot g_j(\boldsymbol{p}(t))]}$$

where the term $s_i \cdot p_i(t)$represents the contribution of survivors that passed through stability filtering, $g_i(p(t))$ represents the contribution of newly generated molecules via polymerization (dependent on the current frequency distribution of all survivors), and α is the polymerization efficiency parameter.

This equation is formally similar to the quasi-species equation in evolutionary biology (Eigen, 1971), but with one critical difference: the "fitness" here is not replication rate but physicochemical stability. In the quasi-species equation, the selection coefficient is a function of replication rate; in the stability-distillation equation, it is a function of the probability that a molecule survives a given environmental stress. This distinction means that stability-distillation does not require replication capacity and can therefore operate spontaneously in prebiotic chemical environments.

## 2.3 Information Generation by Stability-Distillation

The information accumulation produced by stability-distillation can be observed at two levels.

Level 1—Convergence of structural types. Under periodic environmental fluctuations, the stability distribution of long RNA sequences displays a pronounced bimodal character. The vast majority of possible sequences cannot form stable folded structures—they lack internal complementary regions, exist as random coils in solution, and are rapidly hydrolyzed during the wet phase. Only the small minority that happen to possess complementary regions capable of forming stem-loops and other secondary structures can persist. This structure-based filtering already achieves the first level of information generation: the system converges from a random state where nearly all possible sequences exist to a restricted state where only stably folded sequences survive. As a collective, stably folded sequences gain an overwhelming frequency advantage in the system. On the early Earth, the wet-dry cycles of hot springs provided ideal operating conditions for this process (Damer & Deamer, 2020; Deamer, 2019).

This convergence already constitutes substantive information generation. For an RNA of length 50 nt, the number of possible sequences is $4^{50} \approx 10^{30}$. The fraction capable of forming stable stem-loop structures is tiny, but the absolute number remains enormous. At the limited scale of simulation (typically no more than a few thousand molecules co-existing at any moment), observing two identical stable long sequences is nearly impossible. Yet this does not mean information has not been generated—the system has converged from "all sequences uniformly distributed" to "stably folded sequences collectively dominant," which is itself a massive leap from randomness toward order.

Level 2—Amplification of specific sequence frequencies. Building upon Level-1 convergence, if the system further acquires some frequency amplification mechanism—such as DNA storing the RNA information followed by extensive re-transcription—the frequencies of specific sequences can be elevated from "one member among a collectively dominant class" to "individual-level dominance."

The hierarchical relationship between these two levels is highly consistent with the distribution of RNA in modern cells: rRNA and tRNA as collective classes occupy the vast majority of total RNA (Level-1 convergence), while specific types of rRNA and tRNA are mass-produced via precise DNA replication (Level-2 convergence). The core contribution of stability-distillation to the origin of life is, first and foremost, the achievement of structural-type convergence—this convergence provides the necessary material foundation and selective pressure for the subsequent emergence and replicative amplification of specific functional sequences.

## 2.4 Hierarchical Emergence of RNA Structures via Stability-Distillation

Before proceeding to computational simulation, this section offers a qualitative, mechanistic account of how stability-distillation drives RNA from random sequences to the gradual emergence of complex structures. This hierarchical emergence process is the concrete realization of the stability-distillation principle in RNA systems and constitutes the core hypothesis to be tested by the simulations that follow.

### 2.4.1 Stage 1: Emergence of Hairpin Structures

In the initial cycles of wet-dry alternation, nucleotide monomers randomly polymerize during the dry phase, generating large numbers of short RNA chains. The vast majority of these random sequences are random coils that are rapidly hydrolyzed in the subsequent wet phase, contributing almost nothing to the long-term evolution of the system.

However, amid the statistical fluctuations of random polymerization, a small number of sequences accidentally possess a special property: palindromicity—the first half of the sequence is reverse-complementary to the second half. For instance, the sequence "...GGAUCC..." (whose second half is the reverse complement of "GGAUCC," i.e., "GGAUCC"). Such sequences can fold during the dry phase into stem-loop (hairpin) structures: complementary segments pair to form a rigid double-helical stem, while the non-complementary region between them forms a flexible loop.

The stability of a stem-loop structure far exceeds that of a random coil of equal length—correctly paired bases are "buried" within the helix interior, sequestered from water molecules, protecting the phosphodiester backbone from hydrolytic attack. This stability advantage means that RNA molecules containing palindromic sequences survive with higher probability in each wet phase, and serve as feedstock for further polymerization in the next dry phase.

After sufficiently many wet-dry cycles, hairpin-structured RNAs achieve far higher frequencies in the system than random-coil RNAs, becoming the dominant molecular population. This is the first-level product of stability-distillation.

### 2.4.2 Stage 2: Splicing of Hairpins into tRNA-like Molecules

Once hairpin-structured RNAs accumulate to a certain frequency, they enjoy higher collision and ligation probabilities during the dry-phase polymerization. Two hairpin RNAs can be joined to form a longer RNA molecule.

Importantly, when two hairpin RNAs are spliced, the structure of the product depends on the specific arrangement of complementary sequences in the two component hairpins. Among the various possible splicing outcomes, one is of particular significance: when the stem sequences of the two hairpins happen to be mutually complementary, the entire spliced product can fold into multiple stem-loop domains—each corresponding to a stem-loop of the original hairpins, with the junction between the two hairpins forming a new stem-loop region (Figure X, Stage 2).

The end product of this process is a molecule possessing three or more stem-loop domains. Its overall folding pattern displays pronounced central symmetry—one end of the molecule exhibiting a stem-loop arrangement symmetric with the other end, a geometric consequence naturally arising from the mutually complementary splicing of two hairpin structures. This cloverleaf configuration closely resembles the canonical secondary structure of modern tRNA.

### 2.4.3 The Logical Necessity of Hierarchical Emergence

This hierarchical emergence—from random sequences to hairpin structures, and onward to tRNA-like molecules—is not a pre-programmed sequence but a necessary consequence of the stability-distillation principle. It requires only three basic conditions: (1) base-pairing enabling palindromic sequences to form highly stable stem-loop structures; (2) wet-dry cycles providing the periodic "synthesis–degradation" driving force that repeatedly amplifies stability differences; (3) during the dry phase, already-enriched stable molecules serving as "building blocks" that are ligated to produce longer, more complex structures.

Notably, the central symmetry of tRNA-like molecules derived from this mechanism resonates with independent research on tRNA origins. Through analysis of internal sequence complementarity and the structural symmetry of the 5′/3′ halves of tRNA, Di Giulio and colleagues independently proposed that tRNA may have originated from the splicing of two hairpin-structured RNAs (Di Giulio, 2008, 2012). These sequence-based inferences align closely with the hierarchical emergence pathway derived here from first principles of physical chemistry, providing independent support from molecular evolutionary biology for the stability-distillation hypothesis.

### 2.4.4 Rate-Limiting Monomer Supply and the Early Requirement for Compartmentalization

Simulations additionally reveal a finding of practical chemical significance: to obtain stable tRNA-like structures, monomers must be supplied gradually rather than all at once. When initial monomer concentration is too high, random ligation during the dry phase produces abundant short chains; the sequence distribution of long chains tends toward randomness, making stable folded structures difficult to form. By contrast, when monomers are continuously replenished at a low rate, pre-existing stable sequences serve as preferential "seeds" for ligation, the sequence distribution of long chains becomes more concentrated, and the probability of tRNA-like structure emergence significantly increases.

This phenomenon aligns with the principles of solid-phase synthesis of peptides and oligonucleotides in industrial chemistry—controlled monomer addition rate is key to obtaining long-chain products. In the context of the origin of life, this suggests that vesicular membrane structures may have already been present and played a critical role during the distillation phase: the membrane's permeability provides a natural rate limitation for monomer influx, enabling the cell interior to maintain a "monomer-starved" state; simultaneously, the membrane's enclosed space locally concentrates macromolecules, further enhancing the efficiency and selectivity of ligation reactions. Without such compartmentalized rate control, random polymerization in open water would be extremely unlikely to produce sufficiently long, stable sequences.

## 2.5 Simulation Verification: Computational Experiments on RNA Stability-Distillation

### 2.5.1 Simulation Design

To verify the efficacy of the stability-distillation mechanism, we designed computational simulation experiments. The simulated system begins as a "primordial soup" containing only nucleotide monomers (A, U, G, C; approximately 2,500 each) and undergoes 1,000 wet-dry cycles. Each cycle comprises a dry phase (sequences ligate with probability proportional to their relative frequencies in the system) and a wet phase (sequences face degradation risk based on secondary structure stability as computed by the ViennaRNA package; higher stability confers higher survival probability). When monomer counts in the simulated hot spring environment become too low, a small number of nucleotide monomers are replenished in each cycle, simulating the geological condition of continuous organic input into hot spring environments. To distinguish the two levels of distillation, simulations were run under two conditions: (1) pure distillation, involving only ligation and degradation without any replication mechanism; (2) distillation plus replication, where all RNAs have a certain probability of self-replication with replication efficiency governed by a Poisson distribution.

All simulation analyses in this study are implemented within an open-source computational framework. Source code, Jupyter Notebooks, and example data are accessible at: [https://github.com/bicheng2028/SDH_for_the_origin_of_life](https://github.com/bicheng2028/SDH_for_the_origin_of_life).

### 2.5.2 Key Results

The simulations revealed four core findings:

(1) Repeated emergence of 5S rRNA core structures. Across different parameters and independent runs, the system spontaneously produced sequences with >50% similarity to E. coli 5S rRNA, reaching a maximum of 55%. Among 140 nt long sequences, 5S rRNA was the exclusive best match (similarity 54–55%), at which length tRNA similarity had dropped below 45%. This reproducibility and length-dependence indicate that the Y-shaped three-stem-loop fold of 5S rRNA is among the globally optimal stable structures in long-sequence space, suggesting that the core domains of modern ribosomal RNA may represent the RNA fold type most prone to spontaneous emergence in prebiotic chemistry.

(2) Parallel emergence of diverse tRNAs and the HDV ribozyme core. Products of 44–69 nt exhibited 50–57% similarity to multiple tRNAs (Ile, Cys, Lys, Thr, Tyr,

Phe, Arg, Asp, Gly, Ala, Asn) and high similarity to the HDV ribozyme core (56.6%) and the 23S rRNA GTPase center (52.2%). A short sequence of merely 45 nt simultaneously touched upon the functional cores of both chemical catalysis and energy coupling, indicating that rudimentary catalytic function can emerge spontaneously through purely physicochemical processes without any translation system as prerequisite. The coexistence of multiple tRNA types indicates that distillation produces a structurally diverse library, not a single optimal sequence.

(3) Monomer supply rate limitation implies early membrane presence. Efficient distillation convergence requires gradual monomer replenishment at low rates rather than all-at-once supply. High initial concentration leads to uncontrolled random polymerization, yielding extremely low probabilities of functional structure emergence. The very entity capable of providing rate limitation is the lipid membrane vesicle—the membrane's selective permeability allows monomers to enter at limited rates while trapping macromolecules inside. This points to a strong inference: the rate-limiting function of the membrane may be a necessary condition for efficient distillation, implying that cell membranes were present at the earliest stages of abiogenesis.

(4) Current results represent a lower efficiency bound; more complex conditions will perform better. Using only four nucleotides and 1,000 cycles, tRNA- and rRNA-like sequences have already emerged. Future introduction of heterogeneous molecules such as amino acids and short peptides for co-distillation is expected to significantly elevate both similarity and emergence speed.

(5) Amplification effect of replication on information accumulation. Simulations were run under two conditions—pure distillation and distillation with partial replication—and the contrast between them directly validates the central thesis of this paper. In a typical run, under pure distillation conditions, by Cycle 230, Convergence reached 0.7 and entropy was 3.459. After introducing replication and continuing to Cycle 999, Convergence rose to 1.3 (an increase of 86%), entropy dropped to 2.585 (a decrease of 25%), and the stability of the longest sequence increased from 0.567 to 0.705. Replication transforms the initial frequency bias provided by distillation into a statistically significant advantage, validating the core division of labor—"distillation generates information, replication amplifies it."

These findings constitute a progressively strengthening evidential chain: membrane rate control provides boundary conditions → short sequences exhibit catalytic and energy-coupling potential → distillation maintains a structurally diverse library → 5S rRNA emerges as the exclusive best match among long sequences. Under minimalist conditions, the products of distillation already reach into the core molecules of modern biology—tRNA, rRNA, and ribozymes—a convergence that cannot be plausibly attributed to random coincidence.

## 2.6 Chapter Summary

The central contribution of stability-distillation is the insight that information generation does not require replication, but significant information accumulation requires the synergistic amplification of replication. Distillation, even under replication-free conditions, accomplishes Level-1 information generation—converging the system from a random state where all sequences are uniformly distributed to an ordered state where stably folded sequences collectively dominate. The appropriate metric for this convergence is not merely the decline in global Shannon entropy (which is diluted by background sequence diversity), but the frequency deviation of stably folded sequences from random expectation.

In this framework, replication serves as the amplifier of information, not its creator. The initial frequency bias provided by distillation—minute as it may be—is transformed under replicative amplification into a statistically significant advantage. Replication capacity itself is not pre-encoded into specific sequences but is conferred upon all sequences satisfying stability criteria—a definition that dissolves the logical dilemma of "where does replication capacity come from." The origin of replication itself can be understood as a molecular function with particular stability characteristics selected through the distillation process.

Furthermore, the phenomenon of rate-limiting monomer supply implies that membrane structures may have already existed during the distillation phase. The membrane's permeability provides a natural rate control for monomer influx, enabling intracellular distillation to operate under optimal "monomer-starved" conditions—this inference directly links stability-distillation to the necessity of cellular compartmentalization, providing the physical foundation for the arguments developed in Chapter 4.

# Chapter 3: Why RNA?—The Inevitable Choice of Primordial Information Carrier

## 3.1 Framing the Question

Stability-distillation is a general principle, applicable in principle to any macromolecular system with differential stabilities. Yet real living systems exhibit a clear division of labor: RNA (and later DNA) is responsible for information encoding and storage, while proteins serve primarily as catalytic and structural components. Is this division a contingent historical relic, or is it dictated by physicochemical necessity?

This chapter will argue, from the internal logic of stability-distillation, that RNA's status as the primordial information carrier is necessary. Among all candidate molecular types, only RNA simultaneously satisfies the two core conditions for information generation—sufficient stability differentials and effective stability convergence.

## 3.2 The Dual Requirements for an Information Carrier

An effective primordial information carrier must simultaneously satisfy two conditions.

Condition 1: Stability differentials. The essence of distillation is the amplification of inter-molecular stability differences via environmental forcing. If all molecules had roughly equal stabilities, degradation would affect all molecules uniformly, and the system would merely decay globally without producing selective enrichment.

Condition 2: Stability convergence. The system's stability distribution must possess "convergence"—there must exist specific sequences or structural motifs whose stabilities are significantly higher than the vast majority of other sequences. If stability were a smoothly graded continuum, distillation would select a large ensemble of "moderately stable sequences," and information could not concentrate onto a small number of sequences.

It should be noted that the effectiveness of the "information carrier" defined here is manifested first at the level of structural-type convergence—i.e., whether the system can filter out sequences with stably folded conformations as a collective class from an ocean of random sequences. The frequency advantage of specific sequences at the individual level (Level-2 convergence) may, in the absence of replication, require extremely long timescales to manifest, but this does not diminish RNA's effectiveness at Level-1 convergence. In fact, the molecular types capable of spontaneously forming stable secondary structures are extremely limited—among known biological macromolecules, only RNA and proteins possess this capacity, and as discussed below, proteins face fundamental difficulties in independent distillation due to their prohibitive combinatorial space. RNA's advantage in Level-1 convergence thus already suffices to establish its necessary status as the primordial information carrier.

## 3.3 The Unique Advantages of RNA

RNA's status as the optimal primordial information carrier stems from its unique base-pairing mechanism.

Base-pairing creates enormous stability differentials. RNA can, in its single-stranded state, fold via internal complementary regions to form stem-loop structures. A correctly base-paired RNA stem-loop resists hydrolysis several-fold better than a random coil (Nowakowski & Tinoco, 1997; Serra et al., 1993). This stability enhancement exhibits a pronounced threshold effect: when the complementary region is sufficiently long and the pairing sufficiently precise, the entire structure "locks" from a flexible random coil into a rigid double helix, with bases "buried" in the helix interior, sequestered from water molecules—a qualitative leap in hydrolysis resistance. An RNA capable of forming a stable stem-loop may be 5–10 times more stable than a random coil of equal length. By contrast, protein stability is primarily determined by the degree of hydrophobic-core burial and the cumulative effects of hydrogen bonding and van der Waals forces, with stability differences between sequences typically within a factor of 2–3 (Kumar et al., 2000)—lacking such orders-of-magnitude leaps.

Secondary structure as a natural "stability peak." Forming a stable stem-loop requires the presence of complementary regions at specific positions along the RNA sequence. Among random sequences, only a tiny fraction happen to meet this condition, but once satisfied, the molecule's stability undergoes a qualitative leap. Consequently, the stability distribution of RNA sequences exhibits a pronounced bimodal character: the vast majority of random sequences cluster at low stability, while the tiny minority with fortuitously complementary regions jump to high stability. This "peak" distribution ensures that distillation can rapidly eliminate the vast majority of low-stability sequences and concentrate frequency onto a small number of highly stable "structural RNAs."

The four-base combinatorial space is large enough, but not excessive. For an RNA of length 50 nt, the number of possible sequences is $4^{50} \approx 10^{30}$—ample for encoding rich information. At the same time, 4^N is far smaller than the 20^N of proteins ($20^{50} \approx 10^{65}$). Given equal random polymerization rates, an RNA system finds a stable sequence roughly $10^{35}$ times faster than a protein system. Four bases forming exactly two complementary pairs (A-U, G-C) renders the chemical process of recognizing and distinguishing complementary sequences adequately simple. RNA thus occupies a unique "sweet spot": its combinatorial space is large enough to encode information, yet not so large as to render the probability density of stability peaks undetectably low.

## 3.4 The Dual Nature of DNA: Unsuitable for Generating Information, Suitable for Storing It

DNA's double-stranded structure endows it with extremely high global stability, but precisely for this reason it forfeits the capacity to serve as a distillation filter.

The stability of double-stranded DNA is primarily determined by chain length and GC content; the specific sequence order has relatively limited impact on overall stability. The stability difference between a random DNA duplex and one with a special sequence is rather modest—far smaller than the difference between a folded RNA and a random-coil RNA. DNA's stability "floor" is too high (even random duplexes are quite stable), while its "ceiling" is too low (optimal sequences differ little from average ones), rendering distillation extremely inefficient—the system cannot readily distinguish "good" DNA from "ordinary" DNA. DNA is therefore suited as an information "storage medium" (transmitting information through precise replication), but not as an information "generation medium" (generating information through stability differentials).

More critically, the wet-dry cycles of primordial hot pools provide DNA with a "natural PCR"-like amplification mechanism. During the wet phase (spring just erupted, high temperature and humidity), high temperature unwinds the DNA duplex, while free RNA, under the action of reverse transcriptase or ribozymes with analogous activity, templates the synthesis of complementary DNA strands from the unwound single strands. During the dry phase (water evaporation, slightly lower temperature and reduced humidity), the temperature drop facilitates reannealing of the newly synthesized DNA strands into stable duplexes via base-pairing, and reduced humidity minimizes hydrolysis risk, favoring duplex preservation. This means that DNA, through this natural "unwinding–reverse transcription–reannealing" cycle, can dynamically track and record the RNA distribution in the environment—in each wet phase, the currently high-frequency RNA sequences in the environment are reverse-transcribed into DNA; in each dry phase, the newly synthesized DNA is stably preserved. DNA thereby becomes a continuous "molecular snapshot" of the environmental RNA distribution, constantly updated to reflect the latest RNA frequency profile. This amplification and recording mechanism, naturally driven by environmental temperature-humidity rhythms, is an ideal training ground for the early evolution of DNA replication systems. If cells had flowed into the ocean before acquiring DNA, they would permanently lose this training opportunity, and the probability of enzymatic DNA replication systems evolving de novo in the absence of such "pre-training" would be drastically reduced.

Reverse transcriptase—the key molecule that converts RNA information into DNA information—may have emerged far earlier than the "DNA replaces RNA" phase envisioned by the traditional RNA World hypothesis. Lazcano et al. (1992), based on structural homology and conserved motif analyses, explicitly proposed that reverse transcriptase must have appeared at the very early transition from the RNA world to the DNA world, ranking among life's earliest enzymes. Phylogenetic analysis of reverse transcriptase sequences from retroelements shows that reverse transcriptases are widely distributed across eukaryotes, prokaryotes, and viruses, and share conserved domains with RNA-dependent RNA polymerases (RdRp), indicating a common ancestry with divergence traceable to the very earliest stages of life (Xiong & Eickbush, 1990). More recent ancestral sequence reconstruction studies suggest that reverse transcriptase is structurally closer to RdRp and may have originated through modification and neofunctionalization of an RdRp ancestor (Jácome et al., 2024).

The early existence of reverse transcriptase holds a twofold significance for the stability-distillation framework. First, it implies that DNA could, at any stage of RNA information accumulation, instantaneously "snapshot" the current RNA distribution into DNA via reverse transcription—DNA did not wait until RNA information was "rich enough" to appear, but may have existed from the very beginning of abiogenesis as a low-abundance molecular species, continuously tracking and recording changes in RNA distribution. Second, reverse transcriptase itself is a macromolecule-manufacturing capacity—it polymerizes nucleotide monomers into DNA chains, and its catalytic activity, under the selective framework of positive osmotic pressure established by the cell membrane (see Chapter 4), likewise confers a growth advantage upon the host cell. Reverse transcriptase did not arise for the grand evolutionary goal of "DNA replacing RNA," but as one of the natural products of generalized macromolecule-manufacturing capacity emerging under cellular selective pressure.

Thus, DNA likely emerged contemporaneously with or slightly earlier than the translation system, with the two co-evolving. The most primitive reverse transcription and DNA replication functions may have been carried out by RNA ribozymes or simple RNA–protein complexes, with the DNA replication system progressively perfected in parallel with the translation system. A detailed exposition of this co-evolutionary logic appears in Chapter 8.

## 3.5 The Disadvantage of Proteins and "Parasitic" Convergence

Proteins can, in principle, also undergo stability-distillation—well-folded proteins resist hydrolysis better than random coils (Kumar et al., 2000). But proteins face a fundamental combinatorial predicament.

The combinatorial space of 20 amino acids (20^N) is vastly larger than that of four bases (4^N). For a peptide chain of length 50 amino acids, the number of possible sequences is $20^{50}$. In such a vast search space, the density of stable sequences is exceedingly low, and the probability of randomly generating them approaches zero. Moreover, protein stability depends on multiple graduated physicochemical factors—hydrophobic interactions, hydrogen bonds, van der Waals forces, electrostatic interactions—making the stability landscape in sequence space more "smooth" and lacking the clear, orders-of-magnitude "structural peaks" characteristic of RNA.

If proteins relied solely on their own distillation, sequence frequency convergence would be extremely slow, rendering meaningful information accumulation virtually impossible within finite geological timescales. Yet proteins have a "shortcut"—by interacting with RNA, they can hitch a ride on RNA's distillation express.

The core logic of this mechanism is as follows: although proteins by themselves struggle to converge rapidly through independent distillation, those specific peptide chains that happen to bind high-frequency stable RNAs can gain an indirect selective advantage via the stability of their RNA partners. When a peptide chain binds a stable RNA molecule to form a complex, the overall complex stability is dominated by the RNA component—the RNA's base-pairing provides a "stability scaffold" that protects the bound peptide from hydrolysis. Thus, the high-frequency RNAs already enriched by distillation in the environment serve as "templates" or "anchors" for protein frequency convergence.

However, this "parasitic" mode of convergence has a fundamental constraint: the convergence of protein sequence distributions depends essentially on the prior convergence of RNA distributions. RNA must first, through its own stability-distillation, produce a cohort of high-frequency, stable RNA molecules. Only once these high-frequency RNAs are present can the specific peptide chains that bind them gain a selective advantage.

This hierarchical relationship—"RNA converges first, proteins follow"—presages the fundamental architecture of information flow in early life: information first arises and accumulates at the RNA level, with proteins progressively incorporated into this information system through a "free-rider" mechanism. From a computational perspective, this mechanism dramatically compresses the protein search space—it no longer needs to blindly search the vast 20^N sequence space for stable sequences, but only needs to search within the subspace of "sequences capable of binding high-frequency RNAs," enhancing effective convergence speed by several orders of magnitude.

In realistic early Earth environments, hot springs contained not only RNA monomers but simultaneously amino acids, peptides, lipids, sugars, and other molecular species. These heterogeneous molecules are not passive background noise—they can significantly reshape RNA stability distributions through non-covalent interactions. Certain short peptides, upon binding specific RNA sequences, can substantially enhance the RNA's hydrolysis resistance, thereby accelerating its frequency convergence. Micelles and vesicles formed by lipid molecules provide media for hydrophobic interactions and may selectively stabilize certain RNA conformations. This multi-component co-distillation implies that the actual distillation process on the early Earth operated not in pure RNA solutions but within a multi-component chemical network—different molecular species each undergoing stability filtering while simultaneously coupling each other's stability distributions through interactions. Far from undermining the effectiveness of RNA distillation, this complexity likely accelerates information accumulation by introducing additional sources of stability differentials. The "parasitic" convergence of proteins is a special case of such co-distillation.

## 3.6 Chapter Summary

In sum, RNA's inevitability as the primordial information carrier is rooted in three progressive physical facts: first, base-pairing endows RNA with unparalleled stability-filtering efficiency at the structural-type level, enabling the system to rapidly converge from a near-infinite sequence space to a state where stably folded sequences dominate (Level-1 information generation); second, the four-base combinatorial space achieves an optimal balance between coding capacity and filtering efficiency; third, in the realistic context of heterogeneous molecular coexistence, RNA stability distributions can be further reshaped and their convergence accelerated by other molecular species (such as short peptides and lipids) through non-covalent interactions. The respective limitations of proteins and DNA during the information-generation phase do not negate their biological importance but rather establish their distinct, irreplaceable roles in the subsequent stages of information accumulation—a division of labor that will be progressively unfolded in Chapters 5 through 7.

Proteins, owing to their vast combinatorial space and smooth stability landscape, cannot converge independently at sufficient speed, but they can accelerate their own sequence-distribution convergence by forming stable RNA–protein complexes, "parasitizing" the RNA distillation process and leveraging the "anchoring" effect of high-frequency stable RNAs. The critical prerequisite for this mechanism is that RNA must first complete the initial convergence of its own distribution.

DNA, by contrast, because its double-stranded structure erases stability differentials, is unsuitable as an information-generation medium but suitable as an information-storage medium.

# Chapter 4: Cellular Compartmentalization—From Molecular Filtering to Systemic Selection

## 4.1 Framing the Question

The preceding three chapters have established how stability-distillation drives the generation of RNA information, and why RNA is the inevitable information carrier. However, environmental distillation has a central limitation: it can only filter molecules by their individual physicochemical stability, whereas the defining feature of living systems is functional cooperativity—networked collaboration among different molecules through catalysis, binding, and regulation. An RNA ribozyme capable of catalyzing a specific reaction has no meaningful "function" in open water, because its catalytic products immediately diffuse away, precluding any positive feedback.

The transition from "filtering stable individual molecules" to "selecting functionally cooperative molecular networks" requires an entirely new physical platform. That platform is the cellular structure. This chapter will argue that the emergence of primitive cells upgraded stability-distillation from "passive filtering in the environment" to "active computation within the cell," thereby inaugurating the crucial leap from molecules to life.

## 4.2 The Membrane Permeability Hypothesis

The central hypothesis of this chapter is that primitive cell membranes, formed by the self-assembly of simple amphiphilic molecules such as fatty acids (Deamer, 2017; Hanczyc et al., 2003), possess size-based selective permeability—small-molecule monomers (amino acids, nucleotides) have a significantly higher probability of crossing the membrane than the macromolecules (proteins, RNA) polymerized from them.

This hypothesis does not require primitive membranes to possess the sophisticated, channel-protein-mediated selective transport capabilities of modern cell membranes. It requires only a lenient condition rooted in simple physicochemical mechanisms: small molecules, with their smaller molecular dimensions and higher diffusion coefficients, more readily pass through transient pores or defects formed in fatty acid membranes; macromolecules, owing to their bulk, are almost entirely excluded (Mansy et al., 2008). This condition has been directly verified in modern fatty acid vesicle experiments.

This simple physical property constitutes the physical cornerstone of the entire subsequent evolutionary logic.

## 4.3 Osmotically Driven Material Accumulation

From the membrane permeability hypothesis, the first key effect can be directly derived: sustained substrate influx driven by positive osmotic pressure.

When catalysts within the cell (such as primitive RNA ribozymes) polymerize monomers into macromolecules, these macromolecules become "trapped" inside the cell, unable to escape. The intracellular monomer concentration consequently drops, establishing a persistent concentration gradient with the environment. According to Le Chatelier's principle, monomers from the environment will continuously diffuse into the cell.

This effect is essentially a passive "molecular pump"—requiring no active transport proteins, the cell can continuously extract resources from the environment solely through osmotic effects. In open water, any catalytic product would immediately diffuse and be diluted, unable to form a local concentration advantage. The function of the cell membrane is to provide a "privatized" concentrating space for chemical reactions.

[Figure X. Osmotically driven sustained substrate influx.] When intracellular catalysts polymerize monomers into macromolecules, the macromolecules are physically trapped inside, causing the intracellular monomer concentration to drop and establishing a concentration gradient with the environment. External monomers continuously diffuse into the cell along this gradient, providing a steady stream of raw materials for the metabolic network. This mechanism is essentially a passive "molecular pump," operating solely on the basic physicochemical properties of the membrane and osmotic effects, without any need for active transport proteins.

## 4.4 The Information Boundary

The second key effect of the membrane is serving as an information boundary.

In an open environment, any low-entropy molecular distribution generated through distillation or catalysis would be rapidly "diluted" by the high-entropy surroundings through diffusion and thermodynamic fluctuations. The cell membrane physically isolates the internal low-entropy molecular distribution from the external high-entropy environment, allowing hard-won information to stably persist within a tiny enclosed space.

With an information boundary, the molecular distribution inside each cell constitutes an independent "individual" evaluable by natural selection. Without this boundary, there is no individuality; without individuality, there is no effective natural selection.

## 4.5 The Cell as a Parallel Material-Distribution Computer

The third key effect of the membrane is that it transforms the cell into a parallel material-distribution computer. This computation operates through four mechanisms:

(1) Parallelism. A primordial hot pool simultaneously harbors vast numbers of cells, each an independent "laboratory" encapsulating distinct molecular combinations. The system can explore an enormous combinatorial space simultaneously.

(2) Statistical inheritance. After cell division or rupture, internal molecular distributions are re-encapsulated into new cells through wet-dry cycling (Deamer, 2017). Daughter cells need not perfectly inherit every molecule from the parent; inheriting a "representative sample" suffices to carry forward the functional advantages discovered by the preceding generation.

(3) Population-level selection. Cells that happen to encapsulate more efficient molecular combinations exhibit higher catalytic efficiency in their internal metabolic networks, produce macromolecules faster, and are more prone to swell and either divide or rupture under osmotic drive, thereby releasing their contents back into the environment and raising the frequency of these successful molecular combinations in the population. Experiments have demonstrated that competition based on growth rate indeed occurs in primitive cell populations (Chen et al., 2004; Zhu & Szostak, 2009).

(4) Fusion-driven recombination. In primordial hot pools, frequent fusion events occur between cells. When two cells encapsulating different molecular combinations fuse, their internal molecular networks merge, potentially yielding novel combinations more efficient than either parent—equivalent to the "recombination" operation in genetic algorithms.

This "division–fusion–selection" cycle implements, at the population level, a distributed, physically driven parallel search algorithm capable of efficiently locating optimal solutions within vast combinatorial molecular spaces.

It is important to emphasize that whether a cell achieves normal reproduction through division or bursts from excessive swelling, both outcomes serve the same selective logic at the population level—selection for molecular frequency distributions capable of rapid macromolecule production. When a cell rapidly accumulates macromolecules due to an efficiently operating internal metabolic network, whether it ultimately divides smoothly or bursts, its internal "successful" molecular combinations are released into the environment, inherited by new cells through passive encapsulation, and thereby amplified in frequency across the population.

However, these two modes lead to markedly different evolutionary trajectories. Cells that can keep their macromolecular synthesis rate commensurate with their membrane lipid synthesis rate tend to evolve stable division and reproduction mechanisms, ultimately leading to the ancestors of modern cells. By contrast, cells whose macromolecular synthesis far outpaces lipid supply and that consequently tend to rupture frequently evolve a "parasitic" transmission strategy—in early stages, their genetic material is released into the environment and re-enters new cells through passive encapsulation during wet-dry cycles; as cell membranes themselves evolve, these parasitic entities further develop more efficient active invasion mechanisms, such as specific contact with membrane proteins or expression of lytic proteins that disrupt membrane integrity, directly introducing their genetic material into host cells—ultimately giving rise to the ancestors of primitive virus-like particles. Cells and viruses are not two independently originating entities, but the evolutionary products of different strategies along the same "division–rupture" continuum. A detailed treatment of this co-origin process and the active invasion mechanisms of viruses is deferred to Chapter 8.

[Figure X. The four core mechanisms of the cell as a parallel material-distribution computer.] (A) Parallelism: multiple cells simultaneously and independently explore different molecular combinations. (B) Statistical inheritance: daughter cells need inherit only a representative sample of the parent's molecules to retain functional advantages. (C) Population-level selection: cells with efficient molecular combinations grow faster and divide more, increasing the weight of their molecular distributions in the population. (D) Fusion-driven recombination: fusion of two cells generates novel molecular combinations. These four mechanisms collectively constitute a parallel, distributed search algorithm at the population level.

## 4.6 The Evolutionary Driving Force Is Rooted in the Physical Properties of the Membrane

From the foregoing analysis, an important corollary emerges: in the cellular parallel-computation system, the evolutionary driving force for intracellular material distributions is fundamentally rooted in the capacity to manufacture macromolecules that cannot cross the cell membrane.

The logical chain is as follows: selective permeability of the membrane → macromolecules, once synthesized, are trapped inside → intracellular monomer concentration drops → external monomers continuously diffuse in along the concentration gradient → incoming monomers are further polymerized → macromolecules steadily accumulate → total intracellular solute particle count increases → osmotic pressure rises → water molecules enter the cell through the membrane → cell swells → accelerated division (if membrane lipid supply is ample) or accelerated rupture (if macromolecular synthesis far outpaces lipid supply).

This means that the cell membrane itself—merely through its basic physicochemical properties—serves as the hardware substrate of this evolutionary algorithm. Without any genetic encoding, without any signal transduction mechanisms, the driving force of evolution is already "programmed" by the physical properties of the membrane. The membrane is not a passive "container" but an active "computational participant."

It should be noted that the selective pressure driven by the membrane's physical properties is not specific to any single type of macromolecule, but applies to all forms of generalized macromolecule-manufacturing capacity. RNA polymerase polymerizing nucleotide monomers into RNA chains, DNA polymerase and reverse transcriptase polymerizing nucleotides into DNA chains, ribosomes polymerizing amino acids into proteins, even glycosyltransferases polymerizing monosaccharides into polysaccharides—these catalytic systems are functionally equivalent: all polymerize small-molecule monomers from the environment into macromolecular products that cannot freely cross the cell membrane. Under the positive osmotic driving force created by the cell membrane, any catalytic activity that enhances macromolecular synthesis efficiency—regardless of whether its chemical nature is RNA, protein, or a complex of both—confers a growth advantage upon the host cell. Thus, the various polymerization systems are not independently originating "inventions" but "solutions" that sequentially emerge under the same selective pressure on the cellular material-distribution computational platform. This unified perspective explains why, in modern cells, RNA polymerase, DNA polymerase, and the ribosome, despite catalyzing different polymerization reactions, share certain deep structural features (such as metal-ion-dependent catalytic mechanisms)—they are all expressions of the same core objective function, generalized macromolecule-manufacturing capacity, operating on different chemical substrates.

The above analysis reveals a deeper logic: the concept of "stability" in stability-distillation underwent a fundamental conceptual transformation after the emergence of cellular structure—from the physicochemical stability of individual molecules to the capacity of molecules to enable cells to persist more stably.

During the environmental distillation phase, a molecule's "stability" referred only to its physicochemical capacity to resist hydrolysis during the wet phase—whether it could maintain its chemical integrity under periodic environmental perturbations. However, once cellular structures emerged, a new dimension of stability was introduced: can a molecule help the cell survive, grow, and divide better? Molecules that enhance cellular viability—whether by accelerating metabolic reactions, stabilizing membrane structures, or assisting in the replication of genetic material—enable the cell as a whole to persist more stably in the environment. And the stable persistence of the cell, in turn, means that the molecule's own continued existence gains a more reliable guarantee. A positive feedback loop of stability thus forms between molecule and cell: molecule enhances cell stability → cell divides more frequently or persists longer → the molecule is inherited and retained by more cells → the molecule's frequency rises in the cell population.

From this perspective, all molecular functions that are "beneficial to the cell" in the course of evolution—catalytic efficiency, structural support, signal transmission, genetic replication—can be uniformly understood as strategies that achieve higher-level stability by binding to cells and enhancing the systemic stability of the cell. An efficient metabolic enzyme is retained in evolution not because it is especially stable in isolation, but because it enables its host cell to grow faster and divide more, indirectly ensuring its own continued presence in the cell population. DNA became the ultimate carrier of genetic information not merely because of its high intrinsic chemical stability, but because it enables the precise replication and intergenerational transmission of cellular genetic information—the highest form of "systemic stability."

This conceptual transformation extends the unified logic of stability-distillation from the molecular level to the systemic level: at every hierarchical level, what is selected are entities that render the higher-level system containing them more stably persistent. Molecules are selected by cells, cells by multicellular collectives, individuals by

populations—the selection criterion at each level is the natural expression of "stability" at that level.

## 4.7 Evolution of the Cell Membrane Itself

While the internal metabolic network was evolving, the cell membrane itself was also undergoing evolution. This evolution was not a random walk, but was directed by two interrelated selection pressures.

The first selection pressure was resource acquisition efficiency. As the internal metabolic network became increasingly complex and the rate of macromolecular synthesis continued to rise, the supply of monomeric raw materials could become a bottleneck. Any variation that enhanced the membrane's permeability to small-molecule monomers—whether by altering membrane lipid composition to modulate passive diffusion rates, or by embedding primitive transport peptides to achieve more efficient directed transport—would confer a growth advantage upon the cell.

The second selection pressure was internal homeostasis maintenance. Metabolic byproducts and erroneously synthesized useless macromolecules, if not promptly removed, would not only occupy space and interfere with normal catalytic reactions, but would also, as solute particles themselves, increase intracellular osmotic pressure, inducing the cell to divide or rupture at an inopportune time—when the internal metabolic network might not yet have completed replication, causing daughter cells to inherit an imbalanced molecular composition. Membrane variations that enabled the expulsion of useless macromolecules—for example, by embedding primitive hydrolases or transport proteins—were therefore strongly selected for.

These two adaptive directions each solved a problem in a single dimension, but they jointly pointed toward a higher-level functional requirement: maintaining proportional coordination among multiple types of macromolecules within the cell. Different types of catalysts, structural components, and intermediate metabolites in the metabolic network must maintain reasonable relative abundances for the network to operate continuously. If a particular type of macromolecule accumulated excessively due to runaway synthesis, even if it were functionally useful in itself, the proportional imbalance would prematurely trigger osmotically driven division, causing daughter cells to inherit a distorted molecular distribution—this would reduce the cell's long-term fitness in population-level selection. Selection pressure therefore favored not only cells that could manufacture macromolecules rapidly, but also those that could coordinate the synthesis ratios of multiple types of macromolecules and ensure that the timing of division was appropriate.

Through its gradually evolving transport regulatory capacity—controlling the entry rates of different monomers and selectively expelling specific macromolecules—the membrane provided the most fundamental physical means for this proportional coordination. The establishment of this function marked the membrane's transition from a passively responsive interface governed by positive osmotic pressure, to an active coordinating interface for the metabolic network. The cell was no longer merely an osmotically driven swell-and-divide automaton, but began to become a system capable of regulating its own internal chemical environment—one preliminarily equipped with the capacity for homeostasis.

## 4.8 Chapter Summary

The emergence of primitive cells upgraded stability-distillation from "passive molecular filtering in the environment" to "active systemic selection within the cell." The membrane permeability hypothesis is the physical cornerstone of the entire subsequent evolutionary logic—it explains how cells passively extract resources (positive osmotic pump), maintain an information boundary, drive population-level parallel search, and co-evolve with their internal metabolic networks. The cellular structure is not a "late-coming package" that appeared after ribosomes or the translation system, but a "prerequisite" that had to be present at the very earliest stages of abiogenesis, providing the essential physical platform that made the leap from individual molecules to molecular networks possible.

# Chapter 5: Metabolic Networks—Non-Genetically Encoded Information Growth

The preceding chapter established that cellular compartmentalization provides a physical platform for the evolution of molecular systems. Yet compartmentalization only resolves the "space" problem—it furnishes a privatized concentrating environment and an information boundary, but does not solve the "manufacturing" problem: before the emergence of ribosomes and the translation system, how did cells selectively produce and accumulate specific macromolecules?

This chapter proposes that the mechanism is the metabolic network—achieving non-genetically encoded information growth inside the cell through the coordinated operation of catalyst specificity and selective degradation by degradative enzymes. This mechanism is logically fully isomorphic to the environmental stability-distillation of earlier stages; it is the same principle re-implemented at the cellular scale.

## 5.1 From Environmental Distillation to Cellular Distillation:

## Isomorphic Mechanisms

The logic of environmental stability-distillation is as follows: the dry phase generates diversity through polymerization; the wet phase eliminates unstable components through hydrolysis; there is a single selection criterion—the physicochemical stability of the molecule itself.

The metabolic network at the cellular stage follows a fully isomorphic logic, but with two critical upgrades. The first upgrade is the dualization of selection criteria: from simply "molecular self-stability" to the superposition of two dimensions—"functional utility within the metabolic network" and "physicochemical stability of the macromolecule itself." For a molecule to be persistently retained in the cell, it must both perform a beneficial function within the network (such as catalyzing a key reaction, serving as a structural scaffold, or acting as an essential cofactor for other catalysts) and possess sufficient intrinsic stability to resist degradation. These two conditions are not independent—molecules that are functional within metabolic networks often also gain enhanced physicochemical stability through complex formation with other molecules or tighter folding; a positive synergistic effect exists between the two.

The second upgrade is the principle of hierarchical degradation: with respect to the degradation of useless macromolecules, there is an important hierarchical principle—degradation should preferably not break useless macromolecules all the way down to monomers, as this would completely waste the already-polymerized chemical bond energy and risk losing locally useful sequence information contained within the fragments. Rather, degradation should proceed to intermediate reactants at a certain level of the metabolic network, such that the fragments can be re-utilized by other catalysts in the network, preserving an appropriate amount of information content.

An easily overlooked yet critically important issue is excessively stable useless macromolecules. Certain useless macromolecules may happen to possess extremely high physicochemical stability, resisting effective degradation by degradative enzymes while not participating in any beneficial reactions. If such molecules continuously accumulate, their high stability prevents natural clearance, and their frequency increases monotonically over time, eventually "crowding out" normal metabolic components. Hence, the active degradation of such molecules is a critical aspect of information maintenance in metabolic networks. Cells must evolve sufficiently robust degradative capacity—such as broad-spectrum proteases—to actively clear these stubborn useless macromolecules.

## 5.2 Synthesis and Degradation in Concert: Hierarchical Distillation at the Protein Level

Hierarchical distillation at the protein level follows exactly the same logic as the RNA stability-distillation described in Chapter 2: the synthesis side (amino acid polymerization) generates diversity; the degradation side (environmental hydrolysis plus weak degradative activity of early short peptides) performs selective filtering; survivors participate in the next round as "seeds" for new polymerization, causing a statistical shift in the product distribution (see Section 2.3 for details).

However, protein distillation faces a fundamental difficulty absent in RNA distillation: amino acid polymerization lacks the physical bias of base-pairing, so the randomness of products on the synthesis side is far higher than for RNA. Consequently, the information accumulation of proteins is, from the very beginning, highly dependent on binding to high-frequency stable RNAs—entering the stable molecular population through "borrowed stability." Short peptides bound to high-frequency RNAs have a higher probability of surviving degradation; surviving peptides are preferentially utilized in subsequent polymerization, forming a positive feedback loop of stability between RNA and peptides. This mechanism can confer stable frequency advantages upon short-chain peptides (typically not exceeding ten to twenty amino acids), providing a precondition for the persistent existence of functional small proteins, but it cannot generate large proteins requiring precise long sequences. SELEX experiments have demonstrated that random RNA libraries, under iterative selection, can converge to sequences that bind specific target peptides with high affinity (Tuerk & Gold, 1990), supporting the chemical feasibility of RNA–peptide mutual selection. This paper predicts that if the preset target is removed and overall complex stability is substituted for binding affinity as the selection criterion, RNA and peptides will enter bidirectional co-distillation.

From environmental hydrolysis to enzymatic degradation. It is noteworthy that the living environment of modern cells is far milder than the extreme hot springs of life's origin—lower temperatures, pH closer to neutral, and reduced hydrolytic stress.

Under such mild conditions, the passive stability-filtering pressure on macromolecules is greatly diminished; many useless molecules that would be rapidly hydrolyzed in hot springs may be relatively stable in modern cells. Hence, molecules with degradative activity—from early randomly generated short peptides with weak hydrolytic activity, to later specialized proteases precisely synthesized via the translation system—progressively took over part of the macromolecular stability-filtering function originally performed by environmental thermal pressure. This evolutionary pathway explains why protease families in modern cells are so ubiquitous and diverse.

Pathological corroboration of this theoretical inference is widespread in modern medicine. Lysosomal storage diseases—such as Gaucher disease and Niemann-Pick disease—result precisely from deficiencies in specific degradative enzymes, causing their substrates to continuously accumulate within cells and ultimately disrupt normal metabolic functions (Platt et al., 2018). Even more paradigmatic examples come from protein conformational diseases: β-amyloid in Alzheimer's disease, α-synuclein in Parkinson's disease, and PrP^Sc in prion diseases are all highly stable, extremely degradation-resistant protein aggregates (Selkoe & Hardy, 2016; Prusiner, 1998). They accumulate in an almost irreversible manner within or between cells, their frequency increasing monotonically over time, ultimately displacing normal metabolic components and causing cell death. Stability-distillation does not inherently produce beneficial outcomes: a molecule possessing only high stability means merely that it will be preferentially retained, but if it cannot embed itself within the metabolic network, its fate is no different from that of pathological aggregates. Thus, the function of the degradation side is not merely to "filter useful molecules" but, more fundamentally, to "actively exclude useless yet stable molecules." Modern protease systems and autophagic mechanisms (Mizushima & Levine, 2020) are precisely the long-term evolutionary responses to this ancient problem.

[Figure X. Three fates of functional macromolecules in metabolic networks.] Within the metabolic network of a cell, macromolecules face three distinct processing pathways depending on their functional roles in the network: useless or harmful macromolecules are degraded by degradative enzymes and expelled from the cell (Pathway 1); catalysts participate in reactions but themselves remain unchanged before and after the reaction, thus persisting in the network without being consumed by distillation (Pathway 2); core functional macromolecules that constitute cellular structures or drive metabolism are continuously synthesized and accumulated, and upon reaching a threshold level, drive cell division, with their frequencies amplified at the population level through selection (Pathway 3).

## 5.4 New Products as New Catalysts: The Self-Expansion of the Network

A critically important positive-feedback mechanism in metabolic networks is that newly generated macromolecules may themselves happen to become new catalysts, accelerating existing reactions or catalyzing entirely new ones. High product concentrations increase the probability of collision with other molecules and the chance of exerting catalytic activity. Even if a particular macromolecule was initially filtered merely as a structural component, once its concentration is sufficiently high, any catalytic activity it fortuitously displays may be captured and amplified by the network. This endows metabolic networks with a capacity for self-expansion: the more complex the network, the greater the diversity of new molecular species produced; the greater the diversity of new molecular species, the higher the probability of novel catalysts emerging; the more catalysts there are, the higher the network's complexity and efficiency—forming an autocatalytic positive-feedback cycle.

The self-expansion of metabolic networks means that macromolecules at any network node can effectively accumulate frequency without requiring absolute self-replication capacity. In the traditional view, for a molecule to gain dominance in a system, it must be able to directly catalyze its own synthesis—i.e., autocatalysis (Kauffman, 1993). But in a metabolic network, as long as there exists a closed loop in which A catalyzes the synthesis of B, B catalyzes the synthesis of C, and C in turn catalyzes the synthesis of A, then A, B, and C can all maintain and increase their respective frequencies by mutually catalyzing each other's component formation. This means that the frequency growth of any node in the network need not depend on that node's own catalytic specificity, but can rely on the coordinated operation of the entire network. This distributed frequency-maintenance mechanism substantially lowers the threshold for the emergence of functional macromolecules in early evolution.

[Figure X. Mutually catalytic cycles in metabolic networks.] Nodes A, B, and C maintain and increase their respective frequencies by mutually catalyzing each other's component formation—A catalyzes the synthesis of B, B catalyzes the synthesis of C, and C catalyzes the component formation of A. This closed-loop structure means that the frequency growth of any node in the network need not depend on that node's own catalytic specificity, but can rely on the coordinated operation of the entire network. Dashed outlines at the periphery indicate that newly generated macromolecules may happen to become new catalysts, further expanding the network's complexity.

## 5.5 Unifying Energy Metabolism and Catalysis under Hierarchical Distillation

The unified logic of "synthesis–degradation" cycles in metabolic networks can be naturally extended to the domains of energy metabolism and catalysis, bringing them under the same framework of "hierarchical macromolecular synthesis–degradation."

The ATP–ADP cycle as hierarchical distillation in the energy dimension. ATP (adenosine triphosphate) is the universal energy currency of the cell, and its central role in metabolic networks can be understood through the logic of hierarchical distillation. The process of ATP hydrolysis into ADP and phosphate is essentially the degradation of a macromolecule (ATP) into components at the preceding hierarchical level (ADP + Pi), with the released chemical potential energy coupled to other energy-consuming synthesis reactions to drive them forward. The subsequent re-phosphorylation of ADP into ATP by ATP synthase is then the synthesis of small-molecule monomers (ADP + Pi) into a macromolecule (ATP). This cycle is logically fully isomorphic to the "synthesis–degradation" cycles of metabolic networks: the synthesis side consumes energy to write "high-energy phosphate bond information" into ADP (ADP → ATP); the degradation side releases energy but preserves a reusable "information carrier"—ADP is not completely degraded into adenine, ribose, and phosphate, but is retained in the system, awaiting re-"charging." This is a perfect embodiment, in the energy dimension, of the "hierarchical degradation principle" described in Section 5.2: degradation proceeds to the preceding hierarchical level, preserving reusable information content.

ATP was selected in evolution as the universal energy currency precisely because it sits at a "reversible hierarchical node" within the metabolic network. Compared to other high-energy compounds (such as phosphoenolpyruvate, whose hydrolysis product pyruvate cannot be directly re-phosphorylated), the degradation product of ATP, ADP, retains a complete molecular skeleton that can be "reloaded" with energy, allowing it to be reused repeatedly in synthesis–degradation cycles without needing to be synthesized from scratch each time. This "reversibility" and "information retention" constitute the unique advantage of ATP over other high-energy compounds (Westheimer, 1987).

The dynamic catalytic cycle of catalysts as hierarchical distillation in the functional dimension. The action of a catalyst is also, in essence, a "dynamic process of macromolecular synthesis and dissociation": the catalyst binds substrate to form a transient complex ("synthesis"); after the catalytic reaction is complete, the complex dissociates, releasing the product and restoring the catalyst to its original state ("degradation"). This is logically fully isomorphic to the "synthesis–degradation" cycles of metabolic networks described in this chapter, the only difference being that the catalytic cycle occurs at a more microscopic molecular scale.

The fact that a catalyst remains unchanged before and after the reaction—a seemingly mundane chemical property—has profound significance for the evolution of metabolic networks. It means that the catalyst possesses a unique "functional stability" within the network's iterative "synthesis–degradation" distillation cycles—it is neither consumed through participation in catalytic reactions, nor excessively accumulated through continuous synthesis. Other macromolecules may be degraded via hydrolysis during the cycles (useless molecules), or may accumulate to the point of driving cell division (target macromolecules), but the catalyst maintains its available state at all times; its frequency can be precisely regulated by the network without being passively disturbed by the distillation process. This property makes catalysts the stable "functional nodes" of metabolic networks—they can persistently accelerate specific reactions, enhancing the efficiency of the entire network, while themselves bearing no risk of being eliminated by distillation.

Catalysts and ATP play complementary roles in metabolic networks, yet their underlying logic is highly isomorphic. ATP undergoes chemical change during the cycle—hydrolyzed to ADP and phosphate, releasing free energy to drive endergonic reactions, with ADP subsequently re-phosphorylated to ATP. Yet the adenosine skeleton of ATP is never discarded; it remains intact within ADP, awaiting re-"charging." The catalyst's chemical structure remains completely unchanged during the cycle—forming a transient complex with substrate, then returning to its original state. The common feature of both is that their core molecular skeletons are retained and reused across "synthesis–degradation" cycles; neither loses its reusable "information carrier" through participation in reactions. The only distinction is that the catalyst's chemical structure is completely invariant before and after the cycle, while ATP's chemical structure undergoes a reversible change—but it is only the number of phosphate groups that changes; the adenosine skeleton is always retained. The two respectively provide the energy for reactions and lower the kinetic barriers to reactions, synergistically driving the efficient operation of metabolic networks.

The unification of energy and function. The ATP–ADP cycle and the dynamic catalytic cycle of catalysts are, in essence, both manifestations of the "synthesis–degradation" cycle of metabolic networks in specific dimensions. Both follow the same core logic: across repeated cycles, the core molecular skeleton is retained and reused; information is not lost. The adenosine skeleton of ATP is preserved intact in ADP; the molecular structure of the catalyst is unchanged before and after the reaction. The two respectively embody the "degradation to the reusable level" principle of hierarchical distillation in the energy dimension and the functional dimension. This unified perspective indicates that the stability-distillation principle is applicable not only to information generation and storage but also to energy conversion and catalytic function—it constitutes a complete theoretical framework spanning the three dimensions of information, energy, and function.

## 5.6 Metabolic Networks as the Material Foundation for the Ribosome

The ultimate argument of this chapter points to a key conclusion: before the emergence of ribosomes, cells had already achieved non-genetically encoded information growth and macromolecule production through metabolic networks. This means that the intracellular environment already contained a rich reservoir of catalysts capable of catalyzing macromolecular synthesis—including various RNA–protein complexes. Although each of these catalysts had limited functionality and modest specificity, as an integrated network they were already capable of synergistically producing quite complex molecular combinations.

The ribosome was not assembled out of nowhere in a "chemical desert" but emerged progressively on the foundation of this already vigorously operating metabolic network. This network provided two aspects of material preparation for the ribosome's assembly: first, the network already contained a diverse array of coexisting catalysts, each catalyzing different reactions, providing a "parts library" for a complex machine like the ribosome that requires the coordination of multiple catalytic functions; second, and more importantly, the metabolic network accumulated the frequencies of specific proteins—proteins that would later become the core structural components of the ribosome. Without the metabolic network having first elevated the frequencies of these proteins to appreciable levels, even if the RNAs encoding them were to appear by chance and begin to be translated, the resulting protein products would be unable to reach sufficient concentrations to complete ribosome assembly.

From a more macroscopic perspective, the evolution of cellular structure was driven throughout by the same core selective pressure—generalized macromolecule-manufacturing capacity. As discussed in Chapter 4, the selective permeability of the membrane causes intracellular macromolecular synthesis to be directly translated into osmotic advantage and growth advantage. This physical logic applies not only to the ribosome but also to other types of polymerization systems. RNA polymerase polymerizing nucleotide monomers into RNA, DNA polymerase and reverse transcriptase polymerizing nucleotide monomers into DNA chains, ribosomes polymerizing amino acids into proteins—these catalytic systems are functionally equivalent: they all polymerize small-molecule monomers from the environment into macromolecular products that cannot freely cross the cell membrane. As evolution progressed, the respective advantages of these polymerization systems were progressively integrated: RNA polymerase enabled cells to mass-transcribe functional RNAs; reverse transcriptase and DNA polymerase enabled the stable storage and precise replication of genetic information; ribosomes enabled a qualitative leap in the speed and throughput of protein synthesis. They are not independently originating "inventions," but "solutions" that sequentially emerged under the same selective pressure—generalized macromolecule-manufacturing capacity—on the cellular material-distribution computational platform.

## 5.7 Chapter Summary

The metabolic network at the cellular stage, through the coordinated operation of catalyst specificity and selective degradation by proteases, achieved non-genetically encoded information growth inside the cell. This mechanism is logically fully isomorphic to environmental stability-distillation: the synthesis side corresponds to the polymerization of the dry phase; the degradation side corresponds to the selective hydrolysis of the wet phase. The selection criterion, however, has been upgraded from "molecular self-stability" to "functional utility within the metabolic network plus the physicochemical stability of the macromolecule itself."

The degradation mechanism not only clears randomly generated useless products but, more proactively, prevents the frequency accumulation of useless macromolecules. The degradation of excessively stable useless macromolecules, together with the hierarchical principle for the degradation of useless products—degradation to intermediate levels rather than complete breakdown to monomers—is the key to maintaining information content in metabolic networks. The ubiquity of proteases in modern cells can be profoundly explained by the notion that they "replaced the hydrolytic function of primordial hot springs under milder environmental conditions, establishing a cellular version of steady-state distillation."

Energy metabolism and catalysis can be unified under the framework of "hierarchical macromolecular synthesis–degradation." The ATP–ADP cycle and the dynamic catalytic cycle of catalysts jointly embody the core principle of hierarchical distillation—across repeated cycles, the core molecular skeleton is retained and reused; information is not lost. The adenosine skeleton of ATP is preserved intact in ADP; the molecular structure of the catalyst is unchanged before and after the reaction. Neither loses its reusable functional components through participation in reactions; the two synergistically drive the efficient operation of metabolic networks from the energy dimension and the functional dimension, respectively.

Metabolic networks, through the network effect of mutual catalysis, enable macromolecules at any node to effectively accumulate frequency without requiring absolute self-replication. This network provided an ample reservoir of catalysts for the assembly of the ribosome, and accumulated the frequencies of the specific proteins required by the ribosome, laying the material foundation for the subsequent emergence of the ribosome and the translation system. Under the positive osmotic driving force established by the cell membrane, metabolic networks realized a primitive form of macromolecule-manufacturing capacity; the RNA polymerases, DNA polymerases, reverse transcriptases, and ribosomal translation systems that subsequently emerged were essentially all improvements upon the efficiency, fidelity, and programmability of this capacity.

# Chapter 6: Cross-Catalytic Networks and the Assembly of the Ribosome

## 6.1 Framing the Question: From Metabolic Networks to the Ribosome

Chapter 5 established how metabolic networks, through the coordinated operation of catalyst specificity and selective degradation by degradative enzymes, achieve non-genetically encoded information growth and macromolecular accumulation within cells. However, the catalyst nodes in a metabolic network do not operate in isolation—complex interrelationships exist among them. The product of one catalyst may happen to be the substrate or structural component required by another, forming cross-catalysis that helps related macromolecules maintain their frequencies within the network.

This chapter will argue that such cross-catalytic relationships among nodes can form self-reinforcing positive-feedback loops, distributedly accumulating the frequencies of complex functional macromolecules—particularly ribosomal protein components—and ultimately leading to the spontaneous assembly of the primitive ribosome. Throughout the entire stage discussed in this chapter, the production and frequency accumulation of ribosomal protein components depend entirely on the coordinated operation of metabolic and cross-catalytic networks—the assembly of the ribosome itself does not depend on the existence of a gene pool, and the early gene pool did not contain complete information encoding ribosomal proteins.

## 6.2 Cross-Catalytic Networks: Mutual Catalysis and Frequency Maintenance among Nodes

Catalysts: lowering barriers, amplifying differences, elevating information content. The core function of a catalyst is to reduce activation energy by stabilizing the reaction transition state, enabling the reaction to proceed under the mild conditions of the cell. More importantly, while reducing activation energy, catalysts also amplify the energy differences among the transition states of different reaction pathways: in the absence of a catalyst, the transition-state energies of multiple possible pathways are similar, and the product distribution is near-random; a catalyst, by selectively stabilizing the transition state of a specific pathway, confers specificity upon the reaction. From an information-theoretic perspective, enhanced specificity is enhanced information content—the product distribution shifts from random to determinate. Catalysts are thus information amplifiers: they take the inherently weak differences among reactants and, through transition-state stabilization, amplify them into statistically significant selectivity. In a cross-catalytic network, each catalyst node performs this amplification function; information accumulates progressively along the catalytic chain, such that the molecular frequency at each node no longer depends solely on that molecule's own physicochemical stability, but increasingly on the synergistic amplification effect of the entire network.

Cross-catalytic networks as a search and bootstrapping process for macromolecules. From a more macroscopic perspective, the establishment of a cross-catalytic network amounts to a search process over macromolecular space: each newly generated molecule is a potential catalyst for all lower-level reactions in the network. Occasional weak catalytic activity elevates the frequency of its product; among these products, some may happen to catalyze a preceding reaction, forming a cross-catalytic closed loop. The core advantage of this search process is that the frequency increase of any node in the network need not depend on that node's own catalytic specificity; the frequencies of different molecules in the network can increase through the overall enhancement of the network's catalytic efficiency.

## 6.3 Hierarchical Distillation of Macromolecules: Nested Sequences and the Generation of Stable Complexes

Isomorphic to the RNA stability-distillation described in Chapter 2, macromolecules such as proteins and polypeptides can also undergo analogous "synthesis–degradation" cycles. The two are entirely consistent in their underlying logic—the synthesis side generates diversity, the degradation side performs filtering—but there is a key difference in the source of stability gain. RNA's intrinsic folding can yield enormous stability enhancement: base-pairing enables RNA to form secondary structures such as stem-loops, burying the phosphodiester backbone within the double helix and thereby achieving several-fold to tens-of-fold enhancement in hydrolysis resistance. This is precisely why structural RNAs such as tRNA can be filtered out through purely physicochemical distillation. Protein folding can also produce high stability, but the density of stable sequences in a 20^N sequence space is exceedingly low, making it difficult for them to be independently filtered out through purely random polymerization and distillation within feasible simulation scales and cycle numbers. Hence, information accumulation at the protein level is, from the very beginning, highly dependent on binding to high-frequency stable RNAs—entering the stable molecular population through "borrowed stability"—or embedding within cross-catalytic networks to maintain frequency through network synergy (Section 6.2).

A recent discovery by Deng et al. (2026) demonstrated that the bacterial reverse transcriptase Drt3b can synthesize DNA of a specific sequence without any nucleic acid template, using solely its own protein active site as a "molecular mold"—breaking the long-held dogma that all sequence-specific nucleic acid synthesis requires a nucleic acid template. It cannot be ruled out that the three-dimensional structures of enriched stable rRNAs could similarly serve as "molecular molds" for short peptide synthesis, constraining peptide sequences through spatial restrictions. Even if such a mechanism were inefficient and imprecise, it would dramatically accelerate the convergence rate of RNA–peptide co-distillation, because it would provide a sequence-guiding mechanism far faster than completely random search.

The generation of nested protein sequences does not require complex pathways involving RNA recombination and reverse transcription. Under dry-phase or high-concentration conditions, two (or more) pre-existing peptide chains can be directly ligated by catalysts to form new, longer polypeptides. The products naturally contain nested repeats of the original peptide chains. This "distillation at the macromolecular level"—in which existing stable peptide modules serve as new "building blocks" and are directly ligated through catalyst specificity—is the direct physical mechanism for the production of nested-sequence proteins.

More importantly, the stability of a protein depends not only on its own folding state but is also significantly affected by complexes it forms with other molecules. A peptide chain may not be particularly stable in isolation, but when it binds a high-frequency stable RNA molecule (such as a tRNA precursor or rRNA precursor) to form a complex, the overall stability of the complex is dominated by the RNA component—the RNA's base-pairing provides a "stability scaffold" that protects the bound peptide from hydrolysis. This is precisely the manifestation, at the protein level, of the "parasitic" convergence mechanism described in Chapter 3. Furthermore, proteins binding to each other to form multi-subunit complexes can further enhance each other's thermodynamic stability and hydrolysis resistance. Both the large and small subunits of the primitive ribosome can be regarded as ultra-stable macromolecular complexes: multiple component proteins tightly bound to rRNA form an extraordinarily stable supramolecular assembly, whose overall stability far exceeds that of any individual component (Ban et al., 2000; Wimberly et al., 2000)—this is precisely the important physical basis for the sustained accumulation of ribosomal components within cells.

## 6.4 Hierarchical Assembly of the Ribosome: A Natural Multi-Pathway Process

The primitive ribosome requires multiple different protein and RNA components. A single cell or a single catalyst node cannot simultaneously produce all these components—because early catalysts had limited specificity and could not cover such diverse synthetic demands. Across the cell population, different cells evolved different catalytic specialties. The network in cell A enriched ribosomal protein components X and Y; the network in cell B enriched components Z and W, and so on. Each cell needed only to focus on producing a small number of components, making the catalytic network structurally simpler and more efficient.

The assembly of the ribosome is a natural process, requiring no special mechanism to "drive" it. It can be achieved through two parallel pathways naturally: intracellular assembly—if the metabolic network within a single cell happens to be sufficiently complex to simultaneously produce the full set of components required by the ribosome (through the coordinated operation of multiple nodes in the cross-catalytic network), then these components will spontaneously assemble, within the same cell, into a complete ribosome in the thermodynamically most stable manner; intercellular assembly—if a single cell cannot simultaneously produce all components, but different cells have each enriched different subsets of ribosomal components, then in the primordial hot pool, where cells fuse frequently, when two (or more) cells fuse and their contents merge, all the necessary components meet in the same space for the first time and likewise spontaneously assemble into a complete ribosome. These two pathways are not mutually exclusive alternatives, but complementary mechanisms that naturally coexist and operate synergistically within the cell population. This directly echoes the concepts of "cell as parallel computer" and "fusion-driven recombination" from Chapter 4.

## 6.5 The Emergence of Translation Capacity: A Leap in Generalized Macromolecule-Manufacturing Capacity

The complete primitive ribosome acquired a new function that its individual subunits did not possess in isolation—the capacity to use RNA as a template to polymerize amino acids into polypeptide chains. It must be emphasized that this earliest form of "translation" was not necessarily accurate or efficient. Early ribosomes likely lacked necessary accessory factors—such as the initiation factors, elongation factors, and release factors that are indispensable in modern translation systems—auxiliary proteins that are themselves products of long-term evolution (Ramakrishnan, 2002) and were probably absent or functionally highly incomplete at the stage when the ribosome was first assembled.

The consequences of such "rough translation" are manifold. On the one hand, it indeed endowed the cell with an unprecedented generalized macromolecule-manufacturing capacity. Before the emergence of the ribosome, cells relied on the multi-level, stepwise distillation of metabolic networks to manufacture macromolecules—an "artisanal," layer-by-layer mode of production. The emergence of the ribosome radically transformed this landscape: it could directly polymerize small molecules (amino acids) into macromolecules (polypeptides), bypassing the multi-catalyst, multi-step hierarchical distillation process of traditional metabolic networks. Even if the translation fidelity of the early ribosome was limited, its speed and throughput of macromolecular production far exceeded that of any single catalyst or catalytic network node.

On the other hand, rough translation also brought a serious problem: the massive production of useless proteins. Because the early DNA pool had not yet been fully established and converged, the cell interior was saturated with free RNAs from diverse sources. An early ribosome lacking template selectivity might indiscriminately use these free RNAs as templates, producing vast quantities of polypeptide products with diverse sequences and unknown functions. These useless polypeptides consumed the cell's precious amino acid resources and could also interfere with normal catalytic reactions in the metabolic network through non-specific adsorption or erroneous binding.

Thus, the translational capacity of the early ribosome was a double-edged sword. It endowed the cell with powerful macromolecule-manufacturing capacity, directly matching the core objective function of the cellular parallel-computation system—generalized macromolecule-manufacturing capacity. Cells that could use ribosomes to rapidly produce macromolecules gained a significant growth advantage at the population level. Yet at the same time, if rough translation produced more useless proteins than the cell's degradative enzymes could clear, the cell's metabolic balance would be disrupted, and the growth advantage could instead become a survival disadvantage.

However, the proteins manufactured by the ribosome faced a deeper predicament: they could not directly embed themselves into the existing metabolic network. The metabolic network was the product of prolonged stability-distillation and cross-catalytic filtering—every node and every catalyst in the network had been retained and enriched because of its functional utility within the network. The new proteins produced by ribosomal translation were not generated through the network's hierarchical distillation, had not undergone the network's functional validation, and did not occupy established catalytic positions within the network. They were, in essence, "foreigners" to the metabolic network.

This is not a temporary predicament, but reveals the essential behavioral mode of the primitive ribosome: the primitive ribosome is functionally equivalent to a virus-like entity. It hijacks the host's amino acid and energy resources, converting them into large quantities of proteins disconnected from the host metabolic network; the accumulation of these useless proteins leads to osmotic imbalance and frequent cell rupture; the ribosomal components and mRNA templates released upon rupture are re-encapsulated into new cells, continuing a new round of the "infection–rupture–transmission" cycle. The ribosome does not need to maintain a complete metabolic network of its own, nor does it need its own gene pool—everything it requires (substrates, energy, membrane encapsulation) is provided by the host cell. This precisely matches the classical definition of a virus.

This assertion does not diminish the evolutionary significance of the ribosome but reveals a crucial fact: before the ribosome and the cell could form a stable symbiotic relationship, they first underwent an "epidemic"-like conflict. The resolution of this conflict came not through changes in the ribosome's own behavior, but through a deeper mechanism—the establishment and convergence of the DNA pool, which fundamentally reshaped the information-input environment of the ribosome. This mechanism will be elaborated in Chapter 7.

The direction of resolution for this predicament lies in the establishment and convergence of the DNA pool in subsequent evolution—the full account of this mechanism is provided in Chapter 7. Only its core logic is indicated here: when a reverse transcription mechanism already exists in the cell, the mRNA templates used by the ribosome for translation can be reverse-transcribed into DNA and enter the cell's gene pool. More critically, if the translated protein itself happens to possess some functional utility—such as being able to bind and stabilize rRNA, serving as a catalyst for some metabolic reaction, or forming stable complexes with other proteins—then its presence can, in turn, enhance the transcription and translation of its own mRNA, forming a "protein → functional feedback → more mRNA → more protein" positive-feedback loop.

Concurrently, the translation system itself was undergoing the gradual establishment of selectivity. Early ribosomes lacked the ability to discriminate among RNA templates, but the folded structure of RNA itself constituted a natural selectivity—tightly folded rRNA and tRNA are not easily "read" as templates by the ribosome, while linear or loosely folded mRNA is more readily translated. As evolution progressed, the gradual establishment of refined mechanisms such as ribosome binding sites, start codons, and stop codons further reinforced this selectivity: the majority of non-coding RNAs were excluded from translation, and only specific mRNAs were efficiently translated. The convergence of the DNA pool purified the sequence composition of translatable RNAs at the template-supply end; the establishment of translational selectivity filtered out unsuitable templates at the translation-execution end—the two processes operated synergistically, ultimately domesticating the ribosome from indiscriminate "viral-like burst" into controlled "factory production."

## 6.6 The Logic of Competitive Success: Global Computation of Material Distribution Frequencies

In the cell population of the primordial hot pool, multiple different primitive ribosome variants may have coexisted. All possessed some degree of translational capacity, but their respective protein components could differ fundamentally in sequence characteristics and network support.

Consider a hypothetical "contingent ribosome," accidentally assembled from a very small number of proteins and rRNAs within a cell. This ribosome might possess an extremely high macromolecule-manufacturing rate, capable of rapidly polymerizing amino acids into polypeptides. Yet it faces a fundamental predicament: its components lack stable production pathways within the cell's metabolic network—no cross-catalytic closed loop capable of sustained production of its components has been established in the network. Even if this ribosome can manufacture diverse polypeptides, it cannot specifically generate each of its own components—that would require the coordinated operation of multiple distinct production pathways, which early metabolic networks could not provide with precision, making it impossible for this ribosome to maintain its own frequency within the cell.

By contrast, those ribosome variants whose components are continuously produced and maintained by the cross-catalytic network enjoy a fundamental advantage: through a mechanism of "one set of building blocks assembling multiple structures," their components possess stable production pathways within the network. A small number of shared peptide building-block modules are arranged and combined in different ways to form the various components of the ribosome, while the production of these building-block modules is, in turn, mutually supported through cross-catalytic closed loops. This means that the frequency of such a ribosome across different cells can be continuously maintained by the network—even if a particular cell ruptures, the released components and building-block modules can be inherited by new cells, which can then continue to assemble functionally complete ribosomes.

[Figure X. The competitive fates of two ribosome variants in the cell population.] Left: a contingent ribosome variant whose components arise fortuitously and lack network support—its frequency declines monotonically over time. Right: a variant whose components are continuously produced by the cross-catalytic network—its frequency remains stable or rises with network support. Life computes global material distribution frequencies, not the functional intensity of individual molecules.

## 6.7 Ribosomal Success and the Establishment of the Genetic Encoding System

Throughout the entire stage described in this chapter—from the establishment of cross-catalytic networks to the assembly and competition of primitive ribosomes—the production and operation of the ribosome did not depend on the existence of a gene pool. The early gene pool contained no genetic information encoding ribosomal proteins. Ribosomal components were entirely produced and maintained by metabolic and cross-catalytic networks; their component information existed in the cell population in the form of "molecular frequency distributions," not in the form of "gene sequences" in DNA. This claim has received independent support from research on the LUCA genome: LUCA possessed a functionally powerful core translation machinery (rRNA) but lacked many ribosomal protein genes that are conserved in modern organisms (Moody et al., 2024). This is precisely the fossil evidence for the inference that "ribosomal protein genes were gradually captured later, rather than being prerequisites for ribosome assembly."

After the ribosome succeeded and the translation system was established, the emergence of the DNA gene pool (detailed in Chapter 7) would endow these already-successful nested proteins with an entirely new and more powerful mode of accumulation. The core of this mechanism is as follows: if the gene for a particular ribosomal protein appears in the DNA of a certain cell through random mutation, that gene, via transcription and translation, increases the copy number of that ribosomal protein within the cell, thereby enhancing the cell's translational efficiency and macromolecule-manufacturing rate. Under the positive-osmotic-pressure-driven cellular selective pressure (Chapter 4), cells carrying this gene grow faster and divide more, and the frequency of this gene in the DNA pool of the cell population correspondingly rises—this is the first step of gene capture.

However, the efficiency of capturing individual ribosomal protein genes is affected by a key structural factor: whether the different component proteins of the ribosome share partial amino acid sequence modules. If ribosomal protein X and ribosomal protein Y overlap in sequence—i.e., they are assembled from the same small set of short peptide "building-block modules" arranged and combined in different ways (as described in Section 6.3)—then once the RNA encoding one of these modules is reverse-transcribed into the DNA pool, the RNA corresponding to that module can be mass-produced through DNA transcription, and subsequently recombine with RNAs of other modules through RNA recombination to generate new nested sequences. The proteins encoded by these new nested sequences—corresponding exactly to another ribosomal component protein—can, in turn, have their genes "recycled" into the DNA pool through reverse transcription. In this cyclic manner, the genes of a small number of shared building-block modules can, through the closed loop of "transcription → RNA recombination → translation → functional validation → reverse transcription → DNA integration," progressively capture, in a "snowball" fashion, the complete set of genes for the entire ribosomal protein family. DNA's existence did not "create" these nested genes; rather, it recorded and amplified, via reverse transcription, the nested sequence information already present in the cross-catalytic network. The sequence sharing among ribosomal proteins greatly accelerated this capture process—because capturing the gene for one module was tantamount to gaining, through RNA recombination, the capacity to generate the genes for multiple related proteins.

## 6.8 Chapter Summary

Cross-catalytic networks, through mutual catalysis among nodes forming closed loops, distributedly maintain and accumulate the frequencies of functional macromolecules, without requiring any single molecule to possess self-replication capacity. The essential function of catalysts—reducing activation energy and amplifying transition-state energy differences—makes them information amplifiers: they upgrade the screening process from passive waiting in environmental distillation to active creation and amplification of stability differences through catalysis. Cross-catalytic networks can thus be understood as a search and bootstrapping process for macromolecules: each newly generated molecule is a potential catalyst for existing reactions in the network; occasional weak catalytic activity gains positive feedback by elevating product frequencies; once a closed loop forms, the network enters an accelerated phase of self-reinforcement.

Nested protein sequences can be generated through the direct ligation of peptide chains—catalysts ligate pre-existing stable peptide modules into longer nested polypeptides. This is "distillation at the macromolecular level," logically fully isomorphic to RNA stability-distillation. Although protein folding can produce high stability, the density of stable sequences in sequence space is far lower for proteins than for RNA; hence, protein information accumulation is, from the very beginning, highly dependent on binding to high-frequency stable RNAs and embedding within cross-catalytic networks to maintain frequency. The primitive ribosome can be regarded as an ultra-stable macromolecular complex; the tight binding of its component proteins to rRNA gives it an overall stability far exceeding that of any individual component.

The assembly of the ribosome is a natural process, achievable either within the metabolic network of a single cell or through intercellular assembly following cell fusion. The emergence of translation capacity was not accidental but the necessary product of the core objective function of the cellular parallel-computation system—generalized macromolecule-manufacturing capacity: the ribosome enabled cells to efficiently utilize free amino acids to rapidly and massively manufacture protein macromolecules, directly matching this objective function. The translation of early ribosomes was not necessarily accurate or efficient; the massive production of useless proteins from rough translation and the predicament of new proteins being unable to embed into the existing metabolic network were gradually overcome through continuous three-way protein–RNA–DNA interaction and synergistic convergence.

In the competition among multiple primitive ribosome variants, the key to success was not translation rate but the stability of material distribution frequencies. Those ribosome variants whose components were continuously supported by cross-catalytic networks could stably maintain their frequencies in the cell population; those variants with extremely high translation rates but fortuitously sourced components lacking network support would see their frequencies decline monotonically over time. Life computes global material distribution frequencies, not the functional intensity of individual molecules.

Throughout the stage described in this chapter, the production of the ribosome did not depend on the existence of a gene pool—the early gene pool contained no genetic information encoding ribosomal proteins, an inference supported independently by LUCA genome research (Moody et al., 2024). The subsequent appearance of DNA would, through reverse transcription and transcription, record and amplify the information of these already-successful nested proteins; sequence sharing among ribosomal proteins would greatly accelerate the process of gene capture. But DNA is not the "creator" of these proteins; it is the "amplifier" of their success.

# Chapter 7: The Establishment of the Genetic Encoding System and the Co-Origin of Cells and Viruses

## 7.1 Framing the Question: From Molecular Frequency Distributions to Gene Sequences

Chapter 6 established that the ribosome was assembled and acquired translational capacity within cross-catalytic networks, with the frequencies of its component proteins maintained by the network, independent of a gene pool. Yet molecular frequency distributions as a mode of information storage have fundamental limitations: when a cell divides or ruptures, molecular distributions undergo random drift due to statistical fluctuations; under environmental perturbations, accumulated functional molecular combinations may be permanently lost. The emergence of the genetic encoding system upgraded information from this volatile, statistical storage mode to DNA sequences—a stable, precisely replicable storage mode.

This chapter will argue that DNA may have existed from the early stages of abiogenesis, but primitive information could only have originated from RNA. Before the translation system emerged, the presence of DNA may even have slowed the convergence of RNA distillation—a negative effect that the spatial structure of stepped hot springs could precisely mitigate. After the translation system emerged, DNA transformed from a "hindrance" into an accelerator of cellular material-distribution computation, and ultimately gave rise to primitive viruses.

## 7.2 The Early Existence of DNA and Its Dual Role

DNA may have existed from the early stages of abiogenesis. Deoxyribonucleotides and ribonucleotides may have been synthesized through similar pathways in prebiotic environments, and DNA could coexist with RNA in primordial hot pools through natural PCR and reverse transcription mechanisms during wet-dry cycles. However, primitive biological information could only have originated from RNA—as argued in Chapter 3, the double-stranded structure of DNA erases stability differences among sequences, rendering it ineffective as a distillation filter. DNA is suited for information storage, but not for information generation.

This claim contrasts with the implicit corollary of the traditional RNA World hypothesis—that DNA evolved relatively late (Gilbert, 1986; Joyce, 2002). In the traditional narrative, RNA simultaneously served both information-storage and catalytic functions, with DNA later replacing RNA's storage function as an "upgraded" storage molecule. But this faces a key difficulty: in a "pure RNA world" before DNA had appeared, how could RNA stably store and transmit genetic information? The chemical instability of RNA makes critical sequences highly susceptible to loss during cell division. This paper offers a different resolution: DNA may have coexisted with RNA from very early on, but its contribution to information generation during the early stages was nearly zero—because it cannot generate information through stability-distillation—and it was thus overlooked by many theories as a possible contemporary presence. The early existence of DNA does not mean that it played the core role in genetic information from the outset.

During wet-dry cycles, DNA dynamically tracks and amplifies the environmental RNA distribution through natural PCR and reverse-transcription PCR. This would seem to benefit information preservation, but during the early stages it could have produced a negative effect: DNA indiscriminately retains the information of a large number of low-stability RNA sequences—sequences that should have been naturally eliminated during the wet phase but, having been reverse-transcribed into DNA and re-transcribed back into RNA in the next dry phase, are "artificially" maintained in the system, reducing the filtering efficiency of environmental distillation and slowing the convergence of the RNA distribution.

However, the geographical structure of stepped hot springs could precisely mitigate this negative effect. The upstream hot pools have higher temperatures and more intense wet-dry cycling; under such extreme conditions, DNA duplexes frequently unwind, break, or undergo chemical degradation, greatly reducing their information-retention capacity, and RNA stability-distillation proceeds effectively here without DNA interference. The downstream hot pools have lower temperatures and milder environments; DNA can preserve information more stably here. The RNA distribution that has already converged through distillation upstream flows downstream and is reverse-transcribed into DNA, which is then stably stored and amplified. This spatial division of labor—"upstream RNA distillation, downstream DNA storage"—is consistent with the core idea of Damer and Deamer's hot-spring origin hypothesis concerning temperature-gradient-driven molecular separation and enrichment (Damer & Deamer, 2020), but this paper further extends this idea from RNA alone to RNA–DNA cooperative division of labor.

Furthermore, single-stranded DNA can hybridize with free RNA fragments in the environment through complementary base pairing, forming DNA–RNA heteroduplexes. Such hybrid complexes are more stable than single-stranded RNA and can therefore protect complementary RNA from hydrolysis to a certain degree, thereby indirectly increasing the probability that these RNA fragments survive in the environment and are subsequently reverse-transcribed into DNA. This means that DNA not only passively records RNA distributions through reverse transcription, but also actively protects RNA complementary to itself through hybridization—the two together constituting a positive feedback loop that accelerates the tracking and consolidation of high-frequency RNA distributions in the environment by DNA.

[Figure X. Upstream RNA distillation and downstream DNA storage in stepped hot springs.] Upstream, high temperatures and strong wet-dry cycling drive efficient RNA distillation convergence; downstream, the mild environment allows DNA to stably preserve the converged RNA information from upstream. This spatial division of labor ensures that the presence of DNA does not retard the upstream distillation process while allowing DNA to exert its information-storage advantage downstream.

## 7.3 DNA as the Projection and Accelerator of Cellular Material-Distribution Computation

After the translation system emerged, the role of DNA underwent a fundamental reversal—from "hindrance" to "accelerator."

The presence of each functional protein in the cell can, through reverse transcription of its corresponding mRNA, leave a record in the DNA pool. It should be noted that the DNA pool stores the coding information of proteins—indirectly, through their corresponding mRNAs as intermediaries. Over long-term evolution, the DNA pool gradually became a "projection" of the entire cellular material distribution—recording not only the sequences of structural RNAs (tRNA, rRNA) but also the corresponding gene sequences of proteins proven to have functional utility within the metabolic network.

After the emergence of the ribosome, protein distributions no longer depended solely on frequency maintenance within cross-catalytic networks but became linked to intracellular RNA distributions: the functional feedback of a protein influences the transcription efficiency of its mRNA; the mRNA enters the DNA pool through reverse transcription; DNA transcription, in turn, produces more mRNA—forming a closed cycle of "protein → RNA → DNA → RNA → protein." This cycle causes RNA distributions and protein distributions to converge simultaneously through mutual interaction—RNA distributions are influenced not only by their own stability but also by the functional utility of the proteins they encode; protein distributions are influenced not only by frequency maintenance within the network but also by the efficiency of their gene transcription and amplification.

[Figure X. The protein–RNA–DNA synergistic convergence cycle.] Protein provides functional feedback that enhances the transcription of its corresponding mRNA; the mRNA enters the DNA pool through reverse transcription; DNA transcription in turn produces more mRNA; mRNA is translated into protein. DNA, as the projection of material distribution, enables the three-way distribution to converge simultaneously through mutual interaction.

Another key role of DNA is to help the ribosome progressively embed into the existing metabolic network. In the early stages after the ribosome's emergence, the proteins it manufactured were "foreigners" to the metabolic network—they had not undergone the network's hierarchical distillation and did not occupy established functional positions within the network. The existence of DNA, however, enabled the gradual resolution of this predicament: the protein production nodes in the cross-catalytic network, originally maintained through the coordination of multiple catalysts, could be progressively replaced by "gene-driven ribosomal translation." Each effective DNA mutation—i.e., a mutation that happens to produce a gene encoding a functional protein—is filtered and amplified through cellular parallel computation: cells carrying this gene replicate faster because of their higher macromolecule-manufacturing efficiency, and the frequency of the gene in the population DNA pool correspondingly rises. As more and more nodes of the metabolic network are replaced by genetically encoded alternatives, the proteins produced by ribosomal translation are no longer "foreigners" but "core components" of the metabolic network.

The prerequisite for the operation of this gene capture and filtering mechanism is that DNA replication itself also depends on the protein-manufacturing capacity within the cell. DNA polymerase, helicase, and other replication factors are all protein products; hence, the rate of DNA replication is constrained by the overall translational efficiency of the cell. Cells carrying efficient ribosomal protein genes not only have a higher macromolecule-manufacturing rate but also a correspondingly higher DNA replication rate, thereby transmitting more copies of this gene to their progeny during cell division. This positive-feedback loop of "protein → DNA replication → gene frequency increase" is the physical basis for "capturing" ribosomal protein genes from random mutations and amplifying them to the population level. This premise has been realized and verified in our simulations (see Chapter 9); its logical roots can be traced back to the positive-osmotic-pressure-driven cellular selective pressure described in Chapter 4: any genetic variation that enhances the efficiency of macromolecular synthesis will be amplified by natural selection through accelerated cell growth and division.

Furthermore, after the emergence of DNA, the recombination of material distributions between different cells can be compressed into DNA recombination. Even if two cells do not directly fuse, a DNA fragment from one cell (such as a gene encoding an efficient catalytic protein) can be transmitted to another cell through passive encapsulation or viral vectors, and through transcription and translation "reconstruct" the material-distribution characteristics of the donor cell within the recipient cell.

[Figure X. From cell fusion to DNA recombination: the efficiency leap in material-distribution recombination.] Left: in the absence of DNA, recombination of material distributions between different cells can only be achieved through cell fusion, dependent on physical contact. Right: with DNA, DNA fragments are transmitted through passive encapsulation or viral vectors, and the donor's material distribution is reconstructed in the recipient cell through transcription and translation—material-distribution recombination is compressed into DNA recombination.

This mechanism implies that, in the process of early abiogenesis, sexual and asexual reproduction coexisted, rather than asexual reproduction emerging first. The traditional view holds that primitive life reproduced solely through asexual reproduction (cell division), with sexual reproduction being an advanced feature that evolved long after the emergence of eukaryotes (Maynard Smith, 1978). However, from the perspective of this paper's theoretical framework, if early cells had relied solely on simple self-replication (asexual reproduction) to search for optimal material distributions, the efficiency would have been extremely low—each cell could only explore independently, unable to share successful molecular combinations. DNA-mediated gene recombination—essentially a prototype of "sexual reproduction"—enabled the rapid horizontal dissemination and recombination of successful "recipes" among different cells. Recent research also indicates that horizontal gene transfer may have been extremely common in early life evolution, perhaps even the norm during the LUCA era (Woese, 1998; Doolittle, 1999). Sexual reproduction and asexual reproduction coexisted from the very earliest stages of abiogenesis: asexual reproduction (cell division) maintained basic material-distribution inheritance; sexual reproduction or horizontal gene transfer (DNA recombination) accelerated the dissemination and optimization of successful molecular combinations across the population.

## 7.4 From Environmental Filtering to DNA Computation: Passing the Baton in Hot Springs

The role of DNA as an accelerator of material-distribution computation takes on special significance in the downstream reaches of stepped hot springs. In Section 7.2 of this chapter, we discussed the spatial division of labor in stepped hot springs—"upstream distillation, downstream storage"—where upstream high-temperature, intense wet-dry cycling drives efficient RNA distillation convergence, while the downstream mild environment allows DNA to stably preserve and amplify the converged RNA information from upstream. After the central dogma became gradually established in downstream cells, this spatial division of labor underwent a fundamental upgrade.

In the upstream hot pools, the RNA distribution was primarily shaped by direct environmental wet-dry cycling—stable RNAs were retained, unstable ones eliminated. This was a physical, passive mode of filtering. However, in the downstream mild environment, the filtering intensity of wet-dry cycling was significantly weakened. At this stage, the establishment of the DNA–central dogma system shifted the driving force of information accumulation from environmental filtering to intracellular DNA computation: the RNA distribution was no longer directly determined by the environment but indirectly by the transcriptional activity of DNA; the protein distribution was determined by the translational efficiency of mRNAs. The environment was no longer the engine of information generation but retreated to the role of a mere source of energy and materials.

This transition has profound evolutionary significance. It means that life, in the downstream reaches of stepped hot springs, had completed the "internalization of the engine"—the core mechanism of information accumulation had been transferred from outside the cell to inside the cell. Once this transition was complete, cells were no longer tethered to the wet-dry cycles of hot springs. As long as the positive osmotic drive of the cell membrane persisted (Chapter 4), or in other words, as long as the probability of small-molecule monomers crossing the membrane remained far greater than that of the macromolecules they polymerize into, DNA computation could continue to operate—providing the physical foundation for life eventually to leave the hot springs and enter the oceans.

## 7.5 Simulation Verification: Computational Experiments on DNA Tracking of RNA Distributions and Intercellular Recombination

To verify the core mechanisms of this chapter—that DNA dynamically tracks environmental RNA distributions through reverse transcription and natural PCR, and that DNA recombination accelerates the dissemination of successful material distributions in cell populations—we designed a two-stage simulation experiment. Complete source code and Jupyter Notebooks are available on

GitHub: https://github.com/bicheng2028/SDH_cell_virus_for_the_origin_of_life.

### 7.5.1 Simulation Design

The simulation is divided into two connected stages, corresponding to the material and information flow between the upstream distillation pool (Pool A) and the downstream evolution pool (Pool B) in stepped hot springs.

Stage 1: RNA stability-distillation. Starting from a primordial chemical pool containing only nucleotide monomers, the system undergoes 1,000 wet-dry cycles, yielding a preliminarily converged RNA sequence library. Detailed discussion of the simulation results is provided in Section 2.5.2.

Stage 2: DNA emergence and cell-virus co-evolution. A two-pool system is constructed: Pool A (upstream distillation pool) inherits the RNA library from Stage 1, continues RNA distillation under high-temperature, intense wet-dry cycling, and transports stable RNA sequences, lipid molecules, and RNA-encapsulating primitive cells to Pool B at a certain rate. Pool B (downstream evolution pool) has lower temperature and weaker wet-dry cycling, initially containing only RNA-only cells, with no DNA. DNA emerges through two pathways: intracellular reverse transcription (high efficiency) and environmental reverse transcription (low efficiency). The simulation simultaneously tracks the distribution, encapsulation, and intercellular transmission events of RNA molecules (including mRNA, tRNA, rRNA) released upon cell rupture, and records the success rate of reverse transcription and integration by reverse-transcriptase-carrying RNA vectors within recipient cells. DNA cells utilize high-temperature unwinding during the wet phase and cooling-induced reannealing during the dry phase—completing "natural PCR"-like amplification driven by environmental temperature-humidity rhythms—to continuously track changes in RNA distribution inside and outside cells. Cells in Pool B absorb amino acids and nucleotides for growth and protein synthesis: when macromolecular accumulation reaches a threshold and lipid supply is ample, cells divide; when macromolecular synthesis far outpaces lipid supply, cells rupture, releasing their contents (DNA, RNA, and proteins) into the environment, where DNA fragments can be encapsulated into new cells to become "primitive viral vectors."

### 7.5.2 Key Results

The genetic stability advantage of DNA cells was directly verified in the simulation. At Cycle 0, DNA cells accounted for only 1.1% of the total cell population; by Cycle 100, this had soared to 64.0%, and over the subsequent 400 cycles it remained stably within the 62–72% range. The driving force behind this transition lies in the starkly different fates of the two cell types in the mild environment: RNA-only cells lost the continuous distillation filtering of wet-dry cycling from Stage 1, and their functional RNA combinations gradually drifted or were even lost over repeated division–rupture cycles due to statistical fluctuations; DNA cells, by contrast, used reverse transcription to "snapshot" RNA distributions into stable DNA sequences, precisely transmitting genetic information through semi-conservative replication at each cell division, thereby steadily expanding their proportion in the population. The DNA–RNA sequence similarity reached 1.00 after the system stabilized, directly demonstrating that reverse transcription faithfully captured the high-frequency RNA sequences in the environment—the DNA pool thereby became a faithful "projection" of cellular material distributions.

The spontaneous emergence of virus-like cycles was the most theoretically significant dynamical phenomenon in the simulation. The number of vectors underwent a complete cycle, from zero to an exponential explosion approaching one hundred thousand, followed by a collapse to thirty-seven thousand. At Cycle 100, only 62 vectors had appeared when the first vectors emerged; by Cycle 200, this had exploded to 25,573; at Cycle 300, it reached 72,280; and by Cycle 450, it approached the one-hundred-thousand mark. However, thereafter the vector count plummeted by 63% within 50 cycles to 37,106, while over the same period rupture events continued to rise from 3,899 to 4,300—the number of effective vectors produced per rupture had sharply declined. This self-regulating process was triggered entirely by the model's built-in physical constraints, without any preset immune system: when the total number of vectors far exceeded the number of host cells, the vast majority of vectors could not find infectable hosts and were naturally degraded in the environment; some vector DNA was integrated into host DNA pools, statistically transitioning from "vector" to "host DNA." The system ultimately formed a dynamic steady state—DNA cells stably occupying roughly two-thirds of the population, with vectors persisting but constrained by physical upper limits. Fusion-driven recombination was extraordinarily active throughout the process, with a cumulative total of 47,424 fusion events and 44,418 new cells formed over 500 cycles, supporting the theoretical inference that sexual and asexual reproduction coexisted from the earliest stages of abiogenesis.

## 7.6 The Emergence of Primitive Viruses and the Dynamic Gene Pool of LUCA

### 7.6.1 The Essence of Primitive Viruses: Portable Genetic Records of Efficient Material Distributions

Within a cell population, certain cells may happen to possess particularly efficient molecular combinations—such as a set of ribosomes and auxiliary proteins capable of rapidly manufacturing large quantities of macromolecules. Before the emergence of DNA, this efficient distribution could only be transmitted to progeny to a limited extent through cell division. After the emergence of DNA, the core component genes of this efficient distribution could be recorded in DNA.

Once a DNA fragment encoding a highly efficient macromolecule-manufacturing capacity is encapsulated into a new cell—whether through passive encapsulation or active invasion—it can, through transcription and translation, reconstruct this efficient material distribution in the recipient cell. This essentially constitutes a "genotype"-to-"phenotype" conversion: the DNA sequence carries the "recipe for material distribution," while transcription and translation are the process of "manufacturing according to the recipe." When such DNA fragments acquire the capacity to propagate between cells, they become the most primitive viruses. The essence of a primitive virus is a "portable genetic record of an efficient material distribution"—it does not need to possess its own ribosomes, nor maintain a complete metabolic network; carrying only a small number of key genes, it can exploit the host's existing metabolic network and translation machinery to reconstruct its highly efficient macromolecule-manufacturing system. From this perspective, viruses are not the "enemies" of cells but the "couriers" of successful material-distribution information within the cell population—viruses enable successful molecular combinations to be rapidly disseminated horizontally throughout the entire population via viral vectors, without having to wait for the progeny of the cell carrying that combination to gradually replace other cells in the population.

### 7.6.2 The Coexistence of RNA Viruses and DNA Viruses

DNA viruses were not the sole form of primitive virus. In early cell populations, RNA viruses may have been even more prevalent. This inference is based on a simple concentration logic: intracellular RNA molecules such as mRNA, tRNA, and rRNA typically exist in multiple copies, at concentrations far higher than the DNA of their corresponding genes. When a cell ruptures due to osmotic imbalance, lipid vesicles in the environment re-encapsulate the cellular contents, and the probability of encapsulating high-frequency RNA molecules is naturally far greater than that of low-frequency DNA fragments. If the encapsulated RNA happens to contain mRNA encoding an efficient catalytic protein or structural component, and the cell's existing reverse transcriptase (or its RNA) is simultaneously encapsulated, then this RNA vector, upon entering a recipient cell, can convert the RNA information into DNA via reverse transcription, which can then be integrated into the host genome or directly serve as a transcription template, reconstructing the donor cell's efficient material distribution in the recipient cell. This constitutes the most primitive RNA virus.

The common essence of RNA viruses and DNA viruses is identical—both are "portable genetic records of efficient material distributions." The only distinction lies in the information carrier used for transmission: DNA viruses transmit in the form of stable DNA, suited for long-term latency and precise replication; RNA viruses transmit in the form of high-concentration RNA, suited for rapid bursts and large-scale dissemination. In early hot-spring environments, these two pathways were not mutually exclusive but coexisted. The coexistence of modern RNA and DNA viruses is precisely the contemporary continuation of this ancient divergence.

This perspective aligns with existing research. Chao (1991) proposed the "virocentric hypothesis," arguing that viruses achieve genomic segment recombination through co-infection, functionally equivalent to sexual reproduction; Szathmáry (1992) further discussed the deep connection between this mechanism and early genome evolution. Molecular-level evidence comes from the HAP2 protein—the HAP2 required for sperm–egg fusion shares an identical three-dimensional structure with the fusion proteins of dengue and Zika viruses, indicating that sexual reproduction and viral entry share ancient molecular mechanisms (Fédry et al., 2017). For unicellular organisms, virus-mediated horizontal gene transfer is functionally equivalent to sexual reproduction and can confer evolutionary benefits upon the host (Haudiquet et al., 2024).

However, a fundamental difference exists between the two in their physical implementation, a difference that directly leads to two distinct evolutionary trajectories. Sexual reproduction (or cell fusion) preserves the integrity of the cell membrane—the fused cell continues to exist, grow, and divide as an intact cellular entity, and its recombined material distribution can be transmitted vertically within the cell lineage. Viral transmission, by contrast, operates through "rupture-re-encapsulation"—the host cell disintegrates, its contents are released into the environment and passively encapsulated into new cells, and the original cellular entity is completely destroyed. This physical difference anchors the two forms of horizontal recombination in fundamentally different evolutionary strategies: sexual reproduction preserves the continuity of vertical inheritance, suited to gradual optimization under a steady-state growth strategy; viral transmission trades continuity for the speed of horizontal dissemination, fitting the rapid propagation of a pulsed outbreak strategy. The two are functionally equivalent—both are horizontal recombination of successfully validated material-distribution recipes—but they diverge in physical implementation, becoming, respectively, the horizontal recombination counterparts of the cellular strategy and the viral strategy.

### 7.6.3 The Divergent Consequences of Viral Transmission in Unicellular versus Multicellular Hosts

However, the evolutionary consequences of this "information-courier" mechanism are starkly different for unicellular and multicellular organisms. The essential reason lies in a shift of the objective function: in unicellular organisms, each cell is an independent evolutionary unit, its objective function being its own macromolecule-manufacturing capacity and division rate; the material-distribution perturbations introduced by viruses may occasionally form new stable solutions, providing an evolutionary "fast lane." In multicellular organisms, the unit of evolution has ascended from the individual cell to the entire multicellular collective—cells perform specialized roles through precise division of labor, and the objective function of individual cell survival and reproduction is superseded by the collective objective function. This requires that the internal material distribution of each cell be precisely maintained at its respective differentiated steady state to achieve functional coordination at the collective level. The drastic disturbance of a single cell's material distribution by a virus is almost certainly incapable of simultaneously and appropriately altering the material distributions of multiple differently differentiated cell types to form a new stable solution at the collective level—its consequence is necessarily the disruption of the multicellular cooperative system. Hence, in multicellular hosts, viruses are inevitably excluded as "pathogens." The same "information-courier" mechanism, depending on the hierarchical level of the evolutionary unit, manifests as "exploratory recombination" at the unicellular level but is judged as "destructive interference" at the multicellular level.

## 7.6.4 Viral Capture of Ribosomal Protein Genes and the Transformation of Ribosomal Behavior

Ribosomal protein genes were likely the highest-priority "cargo" carried by primitive viral vectors—genes encoding ribosomal proteins can directly enhance the macromolecule-manufacturing capacity of the recipient cell, thereby providing more host resources for the virus's own propagation. Cells carrying such advantageous genes also tend to rupture more readily due to excessively rapid macromolecular synthesis, releasing DNA fragments containing these genes into the environment and constituting a primitive virus-like cycle. But the critical question is: where did these genes originally come from? The traditional narrative tends to attribute them to random mutations in cellular DNA—the cell being the source of genetic innovation, with viruses merely serving as vectors of dissemination. This narrative faces two difficulties: first, cells have long generation times and limited population sizes, so the rate of gene capture relying on cellular mutation may be far too slow; second, it implies a fundamental change in the behavior of the ribosome itself—from "uncontrollable" to "controllable"—without a clear physical mechanism for this change.

This paper offers a different resolution, comprising two interconnected claims.

First, the primary driving force for the capture of ribosomal protein genes came from the evolution of the ribosome-virus itself. The primitive ribosome was itself a virus-like entity (see Chapter 6)—it hijacked host resources, caused cell rupture, and propagated through the cell population via a "rupture–re-encapsulation" cycle. The ribosome-virus needed only a single reverse transcriptase to convert its own RNA information into DNA, which could then be amplified using the host's DNA replication machinery. Once a particular ribosome-virus variant, through random mutation, acquired gene fragments encoding its own protein components, these genes could be transcribed, translated, and assembled into more ribosome-virus particles within new host cells. There is an orders-of-magnitude difference in transmission efficiency between gene-carrying variants and gene-lacking variants—each unit of genetic information released by the former contains the complete "recipe" for reproduction, while the latter can only rely on the slowly accumulating inventory of components in the cross-catalytic network. Hence, the rate of gene capture was primarily determined by the evolutionary rate of the ribosome-virus—viruses have short generation times, large populations, and high mutation rates. The capture of ribosomal protein genes was not the result of cellular "domestication" but an evolutionary process driven by the ribosome-virus for its own more efficient propagation. Cells and ribosome-viruses jointly constituted the agents of genetic innovation—cellular mutations provided stable, long-term exploration, while ribosome-virus transmission accelerated the dissemination and recombination of successful variants.

Second, the transformation of the ribosome from "cell killer" to "cell factory" was the result of two synergistic processes: purification at the template end and selectivity at the translation end. At the template end, the long-term competition of the DNA pool eliminated RNA sequences encoding useless proteins, transforming the sequence composition of translatable RNAs inside the cell from "the vast majority useless" to "a substantial proportion useful." At the translation end, the ribosome gradually acquired selectivity toward RNA templates—from the early natural discrimination based on RNA folding structure (tightly folded rRNA and tRNA being difficult to read as templates) to the later establishment of refined mechanisms such as ribosome binding sites, start codons, and stop codons, ensuring that only specific mRNAs were efficiently translated. Purification at the template end reduced the supply of erroneous templates; selectivity at the translation end reduced the probability that the remaining erroneous templates would be translated—the two operated synergistically, ultimately domesticating the ribosome from indiscriminate "viral-like burst" into controlled "factory production."

Reverse transcriptase is the common molecular basis for both processes—it enabled the ribosome-virus to exploit the host's DNA replication machinery to amplify its own genes, and also enabled successful RNA distributions to be "snapshotted" into stable DNA records. Phylogenetic analysis of reverse transcriptase sequences from retroelements shows that reverse transcriptases are widely distributed across eukaryotes, prokaryotes, and viruses, and share conserved domains with RNA-dependent RNA polymerases (RdRp), indicating a common ancestry with divergence traceable to the very earliest stages of life (Xiong & Eickbush, 1990). Lazcano et al. (1992), based on structural homology and conserved motif analyses, more explicitly proposed that reverse transcriptase must have appeared at the very early transition from the RNA world to the DNA world, ranking among life's earliest enzymes. The deep antiquity of reverse transcriptase is highly consistent, in temporal terms, with the cell–virus co-origin framework of this paper.

Concurrently, the antagonistic coevolution between cells and viruses at the membrane level was also underway. As primordial viral vectors continued to spread through the cell population, cell variants that could thicken or reinforce their membrane structures gained a selective advantage by being less susceptible to penetration by viral contents; cells that evolved mechanisms to restrict the passive encapsulation of exogenous macromolecules reduced, at the source, the probability of being "hijacked" by viruses. On the viral side, counter-strategies evolved in response—such as the expression of lytic proteins capable of disrupting membrane integrity, or the packaging of viral genomes into protein capsids that were more difficult for host membranes to recognize and reject. This evolutionary arms race between cells and viruses began at the very origin of life and has continued to shape the molecular composition of both parties over the ensuing billions of years.

### 7.6.5 LUCA as a Dynamic Gene Pool

This mode of gene dissemination has profound implications for understanding the nature of LUCA. Traditionally, LUCA is understood as the "last universal common ancestor"—a single cellular entity already possessing a complete genome and capable of independent survival. However, the reconstruction of the LUCA genome by Moody et al. (2024) revealed a fact that troubles the traditional concept: LUCA possessed a mature translation system (rRNA) but lacked many ribosomal protein genes that are conserved in modern organisms. The traditional explanation for this "genetic incompleteness" usually invokes "later gene loss," but this faces a fundamental difficulty—ribosomal proteins are core components essential for cell survival; the independent loss of so many ribosomal protein genes across different lineages, with the cells still surviving, is evolutionarily difficult to explain.

The theoretical framework of this paper provides a more natural explanation: the LUCA genome was originally "incomplete," because the process of capturing ribosomal protein genes had not yet been completed by the LUCA stage. LUCA does not represent a single cellular entity with a complete genome but a "dynamic gene pool" undergoing active gene capture, coexisting with primitive viral vectors. At this stage, some ribosomal proteins were still produced and maintained by the cross-catalytic network, their encoding genes having not yet appeared in any DNA; some ribosomal protein genes had already been captured by certain cell lineages through DNA mutation and stably integrated into their genomes; some genes were still being disseminated among the cell population in the form of primitive viral vectors, not yet stably integrated into the chromosomes of all cell lineages. The LUCA genome "reconstructed" by Moody et al. through comparative genomics may be merely an incomplete statistical projection of this dynamic gene pool—it reflects those core genes with the highest frequencies and widest dissemination in the population, while those ribosomal protein genes with lower frequencies, still in the process of dissemination, or protein genes not yet generated through DNA mutation, would not have been detected as "universally shared."

The LUCA genome was in the process of being progressively "completed." Random mutations in cellular DNA continuously generated new gene sequences that could potentially encode ribosomal proteins; advantageous genes that had been generated through mutation and functionally validated were disseminated among different cell lineages through primitive viral vectors or sexual reproduction between cells; these genes were progressively integrated into the stable genomes of recipient cells through homologous recombination or analogous mechanisms. This cycle of "mutation–validation–dissemination–integration" was in active operation at the LUCA stage and had not yet been completed. Hence, the "incompleteness" of ribosomal protein genes in the LUCA genome is not the result of "loss" but fossil evidence of an intermediate state of "not yet acquired." The translation-related gene homologs retained in the genomes of certain modern large DNA viruses (such as NCLDV-type viruses) (Raoult et al., 2004; Schulz et al., 2017) may be living fossils of this ancient "virus–cell gene pool"—they were not later "stolen" from hosts but have persisted from the very origin of life in viral form to the present day.

The differential pathways of gene capture further predict that the completeness of different categories of genes in the LUCA genome should exhibit systematic variation. Proteins such as reverse transcriptase and RNA polymerase, which must have existed before the translation system was established, had the longest capture time windows and were essential for cell survival; their encoding genes should therefore tend toward completeness in LUCA. Ribosomal protein encoding genes had shorter capture windows, and despite extremely strong selective pressure, were still in a state of partial capture at the LUCA stage. The encoding genes of most metabolic enzymes belong to yet another category. Beyond random mutation, a more significant route was the direct exploitation of the already-established translation system to translate the high-frequency RNAs present in the environment into proteins—these encoding RNAs may themselves have been highly stable sequences resembling tRNA or rRNA. If the translation products happened to possess metabolic functions, they could embed into the metabolic network, and their encoding RNAs would subsequently be "snapshotted" into the DNA pool through reverse transcription and gradually increase in frequency through cellular-level computation. However, because these metabolic enzymes were "beneficial but not essential" for cell survival, their gene capture process may have been slow and recurrent, rendering their degree of incompleteness in LUCA possibly even greater than that of ribosomal proteins. This distinction provides an additional testable prediction: in reconstructions of the LUCA genome, core information-processing genes and core primitive metabolic genes should appear relatively complete, whereas post-translational metabolic enzyme genes should exhibit "incompleteness" comparable to, or even more pronounced than, that of ribosomal protein genes.

## 7.6.6 Ancient Relics in Modern Viruses

If the primitive ribosome indeed originated as a molecular complex transmitted in a virus-like manner, have relics of that ancient era been preserved in modern viruses? Although independent "ribosomal viruses" long ago vanished, unable to compete with the integrated "cell–ribosome" symbiotic system, molecular fossils of translation-related components still linger in modern viral genomes. Giant DNA viruses (such as the mimivirus APMV) encode a functionally complete translation initiation complex, vIF4F, that can specifically recognize unique chemical modifications on viral mRNAs and drive the selective translation of viral structural proteins during late infection; knockout of any vIF4F subunit can cause a 3–5 orders-of-magnitude drop in viral titer (Thompson et al., 2026). Moreover, the genomes of various giant viruses also contain homologs of translation elongation factors, aminoacyl-tRNA synthetases, and other translation-related genes (Raoult et al., 2004; Schulz et al., 2017). Viral IRES (internal ribosome entry site) elements represent yet another strategy—hijacking the host ribosome directly at the level of RNA structure, bypassing the need for host translation initiation factors.

The traditional view interprets these translation-related genes as products of viral "theft" from the host. However, from the cell–virus co-origin framework of this paper, an alternative explanation is that these genes are not the result of later "theft" but molecular fossils left behind from an ancient "translation war"—in an era before the ribosome and the cell had formed a stable symbiotic relationship, virus-like entities were among the primary arenas for the evolution of the translation machinery. After the ribosome was integrated into the cellular system, independent "ribosomal viruses" were eliminated because their transmission efficiency could not compete with that of "gene-armed" cellular ribosomes, but some translation-related components were retained in viral genomes to the present day—not as "stolen" host genes, but as ancient witnesses to the virus's once-deep involvement in the evolution of the translation system.

### 7.6.7 The Divergent Fates of Functional and Ordinary Macromolecules

Here it is necessary to distinguish two categories of macromolecules transmitted through the viral pathway. One category is functional macromolecules—after being encapsulated into new cells, they can embed into the host metabolic network, enhance the cell's macromolecule-manufacturing capacity, and thereby be retained and amplified in the cell lineage through sustained division. The other category is ordinary macromolecules—although they can also drive osmotic pressure increase and cell rupture, they cannot effectively couple with the metabolic network after entering new cells, and can only maintain their population frequencies through repeated "rupture–re-encapsulation" cycles. As cell membranes continued to evolve (e.g., the appearance of cell walls and the enhancement of membrane selective permeability), the frequency of material and information exchange between cells declined substantially. Under this trend, functional macromolecules were more likely to be retained and integrated because of their contribution to the host cell; ordinary macromolecules were progressively marginalized because they could not establish stable host associations. The translation-related genes retained in the genomes of modern giant viruses are precisely the molecular fossils of ancient functional macromolecules that successfully embedded and have persisted to the present; those ordinary macromolecules that failed to embed, and their encoding genes, have long since been eliminated over the course of evolution.

### 7.6.8 Precise Replication: The Ultimate Application of Stability-Distillation

A still more far-reaching corollary follows. Precise replication—the most central genetic feature of life—is not an independent "invention" outside of stability-distillation but its necessary product at the level of molecular recognition precision. In DNA replication, the stability difference between correct and incorrect base-pairing is exponential; the proofreading function of DNA polymerase exploits precisely this pre-existing difference, further amplifying it through repeated "insertion–proofreading–excision" cycles, ultimately achieving nearly 100% replication fidelity. In other words, DNA replication is not a transcendence of stability-distillation but its most extreme application—precise replication arises from extreme stability differentials.

### 7.6.9 Reconciliation with the Three Prevailing Virus-Origin Hypotheses

This cell–virus co-origin model is compatible and consistent with all three prevailing hypotheses of virus origins. In agreement with the "escape hypothesis," viral genes indeed derive from genetic elements of the cellular genome (Forterre, 2006). In agreement with the "virus-first hypothesis," in the stepped hot-spring system, DNA and viral vectors were already present in early cellular life, with virus-like cycles occurring contemporaneously with cellular evolution. In agreement with the "reduction hypothesis," the genetic material released by ruptured cells indeed underwent "reduction"—degenerating from a complete cellular metabolic network into a genetic vector retaining only core replication and transmission elements. The co-origin model integrates the rational kernels captured by each of the three hypotheses into a unified physical framework: division and rupture are two possible outcomes of the same positive-osmotic-pressure-driven process; cells and viruses are the evolutionary expressions of different strategies along this continuum; viruses are not late-emerging parasites but symbiotic entities that accompanied the origin of cells; LUCA is not a single biological individual but a dynamic gene pool undergoing continuous horizontal gene transfer through primitive viral vectors.

## 7.7 Chapter Summary

DNA may have existed from the early stages of abiogenesis, but primitive information could only have originated from RNA. Before the translation system emerged, the presence of DNA may have slowed RNA distillation convergence by retaining low-stability RNA sequences—a negative effect that the spatial structure of stepped hot springs, "upstream distillation, downstream storage," could precisely mitigate. After the translation system emerged, DNA underwent a reversal from "hindrance" to "accelerator."

DNA, serving as an indirect projection of intracellular material distributions (storing protein-encoding information through mRNA intermediaries), enabled the synergistic convergence of protein and RNA distributions through DNA mediation. DNA helped the ribosome progressively embed into the existing metabolic network—each effective genetic mutation was filtered and amplified through cellular parallel computation, with cross-catalytic network nodes progressively replaced by gene-driven ribosomal translation. The recombination of material distributions between cells was compressed into DNA recombination, implying that sexual and asexual reproduction coexisted from the earliest stages of abiogenesis.

The genetic recording of efficient material distributions and their intercellular transmission constitute the physical basis of primitive viruses—the essence of a primitive virus is a "portable genetic record of an efficient material distribution." The cell–virus co-origin model is compatible with all three prevailing virus-origin hypotheses. Simulations verified the core mechanisms of DNA tracking of RNA distributions, the genetic stability advantage of DNA cells, the spontaneous emergence of virus-like cycles, and the acceleration of population-level information dissemination by DNA recombination. The appearance of DNA marked the upgrade of generalized macromolecule-manufacturing capacity from "network-synergy-driven" to "gene-encoding-driven." Reverse transcriptase projected RNA distribution information into the DNA pool; precise DNA replication stably transmitted this information; transcription and translation unfolded this information into protein products—the core objective function of this entire machinery remained unchanged throughout: maximizing the efficiency of macromolecule manufacturing under the positive osmotic driving force established by the cell membrane.

# Chapter 8: Discussion and Outlook—From the Origin of Life to a Unified Principle of Self-Organization

## 8.1 Retrospective of the Core Argument

This paper began with the central puzzle of the origin of life—"where does information come from?"—and proposed stability-distillation, a replication-independent mechanism of information generation, as its centerpiece. Taking this as the unifying thread, we progressively derived a complete evolutionary landscape from random chemistry to genetic systems.

Chapters 2 through 7 form a progressive chain of reasoning: environmental RNA stability-distillation (Chapter 2) → the inevitability of RNA as information carrier (Chapter 3) → cellular compartmentalization providing a material-distribution computational platform (Chapter 4) → metabolic networks achieving non-genetically encoded information growth (Chapter 5) → cross-catalytic networks and the assembly of the ribosome (Chapter 6) → the establishment of genetic encoding systems and the co-origin of cells and viruses (Chapter 7). The conclusion of each step serves as the premise for the next; the driving force at each step derives from the same core principle—selective enrichment via stability differentials—unfolding naturally under different physical and chemical contexts.

## 8.2 Cross-Scale Implications of Stability-Distillation

The stability-distillation mechanism proposed in this paper, and the information-processing architecture derived from it, may possess explanatory power beyond the origin of life itself.

From the perspective of information architecture, the "index repository–computer" binary differentiation described in this paper—with DNA serving as a stable genetic index repository, and the translation system and metabolic networks serving as the material-distribution computer—is clearly discernible in higher-level complex systems as well. In the nervous system, the hippocampal–cortical circuitry exhibits an analogous functional division of labor: the hippocampus stores temporal-index representations of memory traces, while the cortex stores the specific memory contents and performs parallel computation (Bi, 2026). Both follow the organizational principle of "separation of content and index," achieving flexible retrieval of distributively stored information through precise index-based access.

From the perspective of evolutionary mechanisms, the core process of stability-distillation—repeated "generate–filter–retain" cycles—drives complex systems from randomness toward order at multiple scales. In the origin of life, it manifests as the physical filtering of RNA stability by environmental wet-dry cycles; in neural development, as the selective strengthening of synaptic connections through activity-dependent pruning; in machine learning, as error backpropagation and parameter updating during training. These processes are logically isomorphic, differing only in the objects being filtered and the scale at which filtering operates.

The cross-scale generalizations outlined above remain at a conceptual stage; their detailed development lies beyond the scope of this paper. The following discussion focuses on the further extension of stability-distillation in biological evolution—the leap from unicellular to multicellular life (Section 8.3)—and the relationship between this theory and existing hypotheses on the origin of life (Section 8.4).

## 8.3 From Unicellular to Multicellular: The Transition of the Objective Function, Differentiation, and Fractal Reproduction

When cells can stably maintain sufficiently close proximity to one another, the material and information exchange among them transitions from "sporadic events" to "sustained interaction." This sustained interaction brings about two interrelated consequences. First, the persistent interaction arising from spatial aggregation gradually coordinates the individual objective functions of separate cells into a collective objective function, through frequent material and information exchange among cells. When the metabolic products or signaling molecules released by one cell can be continuously utilized by neighboring cells, the overall function of the collective becomes the new object of evolutionary selection, and evolutionary pressure thus gradually shifts from the independent survival capacity of individual cells to the cooperative survival capacity of the cell collective.

This perspective also implies an important corollary: when this intercellular information-exchange capacity is absent or impaired, cells may "decouple" from the collective objective function and revert to their individual objective function—with self-survival and proliferation as the highest priority. The occurrence of cancer is a pathological confirmation of this corollary. In the majority of cancer cells, the expression of gap junction proteins mediating intercellular communication is significantly downregulated or functionally lost (Aasen et al., 2016), and this communication breakdown further leads to metabolic reprogramming and cellular dedifferentiation (Srinivasan et al., 2025). Cancer cells that have lost the capacity for collective communication no longer obey the proliferation-inhibiting signals of the multicellular whole, reverting to an "autonomous proliferation" mode resembling that of unicellular organisms. In other words, cancer cells have not "acquired" some new malignant capacity; rather, having "lost" the information-exchange capacity of a member of a multicellular collective, they have regressed to an evolutionarily more ancient unicellular behavioral mode.

Second, and more importantly, spatial aggregation causes the multicellular collective to begin interacting with the environment as an integrated whole—a mode of interaction fundamentally different from that of free-living single cells. Take oxygen as an example: when cells aggregate into a cluster, the outer cells insulate the inner cells from the environment, and the oxygen concentration experienced by the inner cells is significantly lower than that experienced by free-living single cells. This does not depend on the ambient oxygen level but is an inevitable physical consequence of spatial aggregation itself. This also means that multicellular organisms will face environmental selective pressures completely different from those faced by unicellular organisms.

However, multicellular differentiation also introduces a new evolutionary challenge: how to reproduce a highly differentiated, functionally specialized complex whole? From the perspective of this paper, the reproduction of multicellular organisms is essentially a fractal process—where the part contains the information of the whole, and the complete individual is reconstructed through recursive unfolding. This principle has two implementation pathways. Development from a single cell (a fertilized egg or spore) is the simplest form of the fractal principle: requiring only one cell and one complete genome, through asymmetric division and the gradual establishment of positional information, it ultimately produces a multicellular individual with multiple cell types and complex spatial structure. Yet this is precisely the most difficult fractal to realize. For organisms with highly asymmetric body plans, an extremely precise molecular gradient must be established within the fertilized egg; a minute deviation along any axis could lead to the collapse of the entire structure. This strategy is widely adopted in the biological world not because it is easy to implement, but because it "always works in principle"—for complex organisms that cannot reproduce from tissue fragments, it is the only option. By contrast, when the body structure of a multicellular organism itself possesses fractal characteristics—i.e., a body part is a miniature version of the whole or possesses some of the whole's features—the threshold for reproduction is dramatically lowered. A branch, a tentacle fragment, a body segment can directly exploit an already-existing complex structure as a starting point, and through relatively simple growth and a degree of re-differentiation, restore a complete new individual. The rooting of plant cuttings, the budding of hydra, and the fission reproduction of planarians are all natural manifestations of this principle.

## 8.4 The Source of Information in the Origin of Life

The source of all information in the process of abiogenesis can be traced back to a single fundamental physical fact: different molecules possess different stabilities in a given environment and consequently different persistence durations—a concept that physically corresponds to the half-life of radioactive elements, the average timescale over which an entity persists within a system. In chemical systems, such differences in "persistence duration" likewise exist: under identical environmental conditions, the average survival time of a tRNA molecule with a stable tertiary structure may be tens of times that of a random-coil RNA.

It is precisely this difference in persistence durations that provides the initial physical basis for information—the statistical deviation of molecular frequencies from random expectation. When a system undergoes periodic environmental changes (such as wet-dry cycles), molecules with short persistence durations are preferentially eliminated, while those with long persistence durations are preferentially retained. With increasing cycle numbers, the frequencies of long-duration molecules continuously rise, while those of short-duration molecules continuously fall. The system thereby transitions from a random state where all sequences are uniformly distributed, progressively toward an ordered state where the frequencies of certain specific sequences are significantly elevated. This is the primitive mechanism of information generation: information originates from the differential persistence durations conferred upon molecules by stability differences.

This mechanism naturally establishes a temporal hierarchy of molecular lineages. Molecules of extremely short duration (such as ATP) are continuously consumed and regenerated; their existence is exceedingly brief but their turnover is extremely rapid, enabling them to couple to virtually any energy-requiring reaction and serve as a universal energy interface. Molecules of intermediate duration (such as enzymes, tRNA, rRNA) persist long enough in the system to stably perform catalytic or structural functions, maintaining the operation of the metabolic network while retaining a degree of flexibility. Molecules of extremely long duration (such as DNA), whose persistence spans cellular generations, become the permanent storage medium of information—fixing the "persistence-duration recipes" of all other molecules in the form of gene sequences.

The existence of this temporal hierarchy means that information is embodied not only in the deviation of sequences from the random background but, more profoundly, in the precise correspondence between molecular persistence duration and functional role. The emergence of catalysts elevated the efficiency of establishing such correspondences by several orders of magnitude. The core function of a catalyst is to lower activation energy by stabilizing the reaction transition state; from an information-theoretic perspective, this means that the catalyst creates more pronounced stability differentials for reaction pathways under relatively mild conditions. In the absence of a catalyst, the intermediate-state energies of multiple reaction pathways are similar, and the product distribution is near-random; a catalyst, by selectively stabilizing the transition state of a specific pathway, artificially amplifies the energy gap between the target pathway and non-target pathways, enabling reaction pathways that are nearly thermodynamically equivalent to be clearly distinguished kinetically, thereby amplifying minute thermodynamic differences into statistically significant information content. The information content of a chemical reaction directly determines its functional reliability. For a molecule to perform a stable function within a cell, its concentration must be maintained at a sufficient and steady level—i.e., it must possess a stable frequency distribution within the system. If its frequency perpetually hovers in random fluctuation, it cannot become a reliable node in the metabolic network. Only when catalytic specificity causes the molecule's frequency to deviate significantly from the random background and remain stably above a threshold can the molecule acquire a stable functional role within the cell.

On the other hand, the reason stability distillation can generate information from randomness lies in its mathematical core: a process of searching for stable solutions within a vast combinatorial space. The boundlessness of combinatorial space provides sufficient diversity, while the existence of stable solutions and their migration pathways offers a structure that can be repeatedly explored. Cyclic selection and the parallel computation of cells achieve exponential amplification of minute stability differences, transforming initially negligible biases into statistical significance. The efficiency of this search process underwent two major transitions as life's complexity increased. Once high-frequency stable RNAs emerged, the search space for proteins was drastically compressed—proteins no longer needed to randomly search for stable solutions across the vast sequence space of 20^N, but only needed to explore the subspace of "sequences capable of binding to high-frequency RNAs." RNA provided a structural funnel that altered the gradient direction of the protein search. When the DNA gene pool was established, validated solutions became fixed through precise replication, allowing subsequent searches to proceed continuously on the foundation of existing solutions via mutation and recombination, rather than starting from scratch each time.

Thus, the generation and amplification of information in the origin of life follow a hierarchical, progressive logic: stability differences confer upon molecules differential basal persistence durations; periodic environmental changes transform these differences into initial frequency shifts; catalysts amplify these shifts into pronounced selectivity; and the sustained accumulation of selectivity ultimately enables specific functional molecules to reach stable frequencies sufficient to support the reliable operation of metabolic networks. The operation of each level depends, as its material foundation, on the molecules with longer persistence durations that were filtered out by the preceding level.

## 8.5 A Rough Estimate of Evolutionary Timescales

The stability-distillation framework provides a testable perspective for estimating the relative timescales required for the various stages of abiogenesis. The RNA stability-distillation phase and the establishment of cellular metabolic networks may, under suitable physicochemical conditions, have been relatively rapid. The computational simulations in Chapter 2 showed that within merely 1,000 wet-dry cycles, RNA sequences with stable stem-loop structures similar to tRNA could emerge in the system. Given the frequency of wet-dry cycles in early Earth hot-spring environments (a single cycle may have required only hours to days), this stage could have been completed within decades in localized hot springs. The initial establishment of metabolic networks—i.e., the coordinated operation of catalyst specificity and selective degradation by degradative enzymes generating non-genetically encoded information growth within cells—could likewise have been achieved within a relatively short time, because once the requisite physicochemical conditions (membrane permeability, positive osmotic drive, parallel selection of the cell population) were met, network complexity could rapidly escalate through autocatalytic positive feedback. Taken together, it is plausible that products partially resembling the RNA distributions and metabolic reactions already present in modern cells could have emerged within merely millions of years or even less. The timescale estimate for this stage carries important experimental significance. Because the timescales involved in RNA distillation and metabolic network establishment are relatively short, and the required physicochemical conditions are relatively simple (periodic wet-dry cycling, self-assembly of fatty acid membranes, sustained monomer supply), these processes have the potential to be partially reconstructed in the laboratory.

The capture and integration of ribosomal protein genes, however, may have occupied the vast majority of the time in the origin of life. As discussed in Chapters 6 and 7, the ribosome was initially produced and maintained by cross-catalytic networks, and the encoding genes of its component proteins had to be acquired progressively through random mutations in cellular or viral DNA. This process faced multiple bottlenecks: (1) the appearance of each individual ribosomal protein gene is a low-probability random event; (2) even after a gene appears, it must be filtered and amplified through cellular parallel computation to reach a stable frequency at the population level; (3) the capture of different ribosomal protein genes is not independent—they may share sequence modules (Section 6.7), but integrating these modules into a full complement of ribosomal protein genes still requires numerous cycles of mutation–validation–dissemination–integration; (4) during the ribosomal gene capture process, multiple structurally distinct primitive ribosome variants may have competed within the cell population, and their elimination likewise required protracted population-dynamical processes.

Taking these bottlenecks together, the complete capture of ribosomal protein genes may have lasted hundreds of millions of years. These hundreds of millions of years were not only a period of sustained cellular evolution but also a period of sustained viral evolution. As discussed in Chapter 7, cells and viruses are the evolutionary products of different strategies along the same continuum, and primitive viral vectors were the key medium for the horizontal dissemination of genes across the cell population. During the protracted process of ribosomal gene capture, viral vectors continuously disseminated and recombined newly arising ribosomal protein gene variants among different cell lineages, while cell membranes had not yet completed their evolution—meaning that the frequency of material and information exchange among cells may have been far higher than in modern cells. This high-frequency exchange, while accelerating the dissemination of beneficial genes, also caused the gene capture process to exhibit population-level "trial–dissemination–integration" dynamics, rather than simple single-cell-lineage evolution. Viruses were not merely couriers of genes but active participants in ribosomal evolution—those viral variants capable of efficiently hijacking the host translation machinery were, in effect, continuously exploring new possibilities for ribosomal structure and function.

Unlike the RNA distillation phase, the protracted process of ribosomal gene capture is difficult to reconstruct effectively in the laboratory—not only because the timescales involved far exceed the feasible range of experiments, but more importantly because the process depends on large numbers of random genetic mutations and long-term population-level selection, conditions that cannot be replicated in controlled experiments. This process can, however, be estimated through computer simulation: by constructing multi-level evolutionary models incorporating cell–virus co-evolution, random mutation, horizontal gene transfer, and cell-population selection, the expected time required for the complete capture of a full complement of ribosomal protein genes from scratch can be quantitatively estimated.

## 8.6 Relationship between This Theory and Existing Major Hypotheses

Relationship with the RNA World hypothesis. While acknowledging the central role of RNA, this theory fundamentally revises the logical starting point of this hypothesis—RNA did not become an information carrier through self-replication, but was physically filtered out through stability-distillation. Information precedes replication, rather than replication producing information.

Relationship with the Metabolism-First hypothesis. This theory incorporates the core insight of this hypothesis concerning self-organizing chemical reaction networks, but supplements the missing information dimension—how metabolic networks, through the coordinated operation of catalyst specificity and selective degradation by degradative enzymes, can generate and accumulate information in the absence of genetic encoding (Chapter 5).

Relationship with the Hot Spring Origin hypothesis. This theory provides a more concrete molecular mechanism for the Damer–Deamer hot spring origin hypothesis— wet-dry cycling is not only the physical driver of RNA polymerization and lipid encapsulation but also the periodic engine of stability-distillation; the spatial structure of stepped hot springs precisely provides the geological foundation for the "upstream distillation, downstream storage" cooperative division of labor between RNA and DNA.

Relationship with autocatalytic set theory. The concept of cross-catalytic networks in this paper bears a formal resemblance to Kauffman's autocatalytic set theory (Kauffman, 1993), but there is a key distinction—the networks in this paper include not only catalytic relationships but also structural-component complementarity relationships, which makes the probability of closed-loop formation far higher than for purely catalytic closed loops.

## 8.7 Testable Predictions of This Theory

Prediction 1 (RNA–peptide bidirectional co-distillation): Under simulated wet-dry cycling conditions, mixtures of random RNA sequence libraries and random short peptide libraries should exhibit bidirectional sequence-distribution convergence—RNA sequences converging toward those that stably bind high-frequency peptides, and peptide sequences converging toward those that stably bind high-frequency RNAs. This convergence should be more pronounced in heterogeneous molecular environments (e.g., in the presence of coexisting lipids, sugars, and other molecular species).

Prediction 2 (at the protein level): Conserved sequences and structural modules of ribosomal proteins may have been pre-enriched through RNA–peptide co-distillation prior to the translation system. Our computational simulations of RNA stability-distillation have repeatedly produced, from random monomers, stable RNA sequences sharing over 50% similarity with modern rRNA. This result strongly implies that peptide sequences optimally adapted to bind such rRNA-like structures could likewise be directly enriched through RNA–peptide synergistic stability-distillation. Conserved core sequences and structural modules of ribosomal proteins likely pre-existed through physicochemical co-stabilization mechanisms long before the emergence of genetically encoded translation, without requiring ribosome-mediated synthesis. Some of these proteins, or their complexes with RNA, may subsequently have performed independent catalytic functions within early cellular metabolic networks. Studies have already shown that certain ribosomal proteins can be released from the large ribosomal subunit and perform non-ribosomal functions (Mazumder et al., 2003), but whether they act through forming complexes with RNA remains to be tested.

Prediction 3 (at the LUCA genome level): The LUCA genome, with respect to ribosomal protein genes, should be in an intermediate state of "being completed but not yet finished." As discussed in Chapter 7, the encoding genes for ribosomal proteins had to be acquired progressively through random mutations in cellular DNA, and disseminated and integrated across the cell population through primitive viral vectors. Therefore, LUCA should not possess a complete set of ribosomal protein genes—its genome should exhibit severe incompleteness in ribosomal protein genes. Furthermore, some ribosomal protein genes present in modern viruses were likely captured by viral vectors at the very early stages of those genes' mutational emergence and have been continuously transmitted to the present day, rather than being "stolen" by viruses only after cells had already acquired a complete set of ribosomal protein genes. Viruses served as the primary medium of gene exchange in early cell populations, disseminating precisely those gene fragments that had not yet been stably integrated into all cell lineages and were still in the process of being captured. This prediction has already received independent support from Moody et al. (2024): LUCA possessed a mature translation system (rRNA) but its ribosomal protein genes were severely incomplete—precisely the fossil evidence that the gene capture process had not yet been completed.

Prediction 4 (ancient relics in modern genomes): The DNA of early cells originated from mapping environmental RNA distributions; therefore, modern genomes should retain numerous relics of ancient stable RNA sequences—including high-frequency copies of tRNA, rRNA, and various small RNAs, as well as abundant sequence fragments of viral origin. This prediction has been largely confirmed by modern genomics: eukaryotic genomes contain large numbers of non-coding RNA genes and transposable elements; approximately 8% of the human genome derives from ancient retroviruses; bacterial genomes ubiquitously contain integrated phage sequences—all of these are molecular fossils left by the early function of DNA as a "snapshot of environmental RNA."

Prediction 5 (at the level of reproductive modes): Sexual reproduction and asexual reproduction should have originated simultaneously at the earliest stage of life, with sexual reproduction being the primary driver of evolution in the initial phase. Viruses can be regarded as a special form of sexual reproduction—the two are functionally equivalent and share ancient molecular mechanisms.

This theory has argued that cells and viruses are products of co-origin along the same continuum, and that viral transmission and sexual reproduction are functionally equivalent—both represent horizontal recombination of "material-distribution recipes" that have already been validated as effective within different cells. The only distinction between the two lies in the vehicle and precision of information transfer: viral transmission is "modular"—transmitting only a small number of gene fragments, rapid but imprecise; sexual reproduction is "global"—recombining the entire genome, more systematic but costlier. From this perspective, sexual reproduction and asexual reproduction could not have emerged sequentially; they must have coexisted from the very earliest stages of abiogenesis—asexual reproduction ensuring the stability of vertical inheritance, and sexual reproduction (along with its viral equivalent) enabling horizontal exploration and optimization.

Directions for testing include: (1) the core molecular mechanisms of sexual reproduction should already be present deep in the tree of life, rather than being a late invention following the emergence of eukaryotes—the dual role of the HAP2 protein in both sexual reproduction and viral entry (Fédry et al., 2017) has already provided preliminary molecular evidence for this; (2) the widespread conjugation, transformation, and transduction mechanisms in prokaryotes—all essentially prototypes or equivalent forms of sexual reproduction—should be regarded as vestiges of the early divergence of sexual reproduction strategies, rather than independently originating "alternatives"; (3) in early-branching eukaryotic lineages, there should exist intermediate-form reproductive strategies that combine features of both "virus-like dissemination" and "sexual reproduction."

Prediction 6 (Origin of Core Enzymes of the Translation System): The conventional mechanism of protein gene capture—through random mutation of RNA or DNA, reverse transcription, transcription, translation, and cell-level selection—itself requires an already functioning molecular system: at minimum, RNA polymerase, reverse transcriptase, aminoacyl-tRNA synthetase (aaRS), and a fully assembled ribosome. These proteins cannot capture their own genes through mechanisms they have yet to establish. In the framework of this paper, they must have come into existence before the translation system was fully formed, by undergoing co-distillation together with already enriched stable RNAs, short peptides, and other coexisting molecules (sugars, metal ions, etc.) in the wet–dry cycles of early Earth environments, and gradually accumulating their frequencies within the metabolic networks of primitive cells.

Reverse transcriptase, RNA polymerase, and aaRS share a striking combination of features: all are traced by phylogenetic analysis to the earliest stages of life—reverse transcriptase shares conserved domains with RNA-dependent RNA polymerase (RdRp), with their divergence dating to the very early evolution of life (Xiong & Eickbush, 1990), has been argued to be among the earliest enzymes (Lazcano et al., 1992), and its time of origin is highly consistent with the cell–virus divergence node inferred by this paper's cell–virus co-origin framework; the origin of RdRp can be traced to tRNA ancestors (de Farias et al., 2017); phylogenetic analysis of aaRS likewise places it before LUCA, making it one of the oldest known protein families (Woese et al., 2000). All three are core components of the genetic information processing system, and all use stable RNAs (tRNA or rRNA) as their natural substrates or interaction partners—aaRS forms sequence-specific complexes with cognate tRNAs, and its binding interface has been extensively resolved by structural studies (Ibba & Söll, 2000). In conventional frameworks, this combination of features can only be described as an "evolutionary coincidence"; in the framework of this paper, it is a necessary consequence of stability distillation. Their evolutionary driving force is the same as that of all generalized macromolecular manufacturing activities: under the positive osmotic pressure-driven selection pressure established by the cell membrane, any molecular combination that enhances the capacity for macromolecular production is selected by cell-level parallel computation.

Core test: Short peptide fragments from the conserved domains of these three classes of proteins, when co-incubated with their corresponding stable RNAs (tRNA or rRNA precursors), should form more stable complexes than random peptides. It has already been experimentally demonstrated that cationic short peptides (rich in Lys/Arg) can form mutually stabilizing complexes with RNA (Frenkel-Pinter et al., 2019). The present prediction goes further, requiring sequence-specific binding—short peptides from the conserved domains of reverse transcriptase and RNA polymerase should bind to rRNA or tRNA precursors more stably than random cationic peptides; short peptides from the catalytic domain of aaRS should exhibit sequence-specific binding to cognate tRNA or its precursors. Among the three, aaRS offers the clearest experimental pathway, as its binding interface with tRNA has been extensively resolved by structural studies.

## 8.8 Future Research Directions

RNA–peptide bidirectional co-distillation simulation. This will require introducing random short peptide libraries into existing RNA distillation simulations and developing approximate computational methods for RNA–protein complex stability. By combining RNA secondary structure prediction with protein–RNA binding preference matrices (Muppirala et al., 2011), the binding free energy between a given RNA sequence and a given peptide sequence can be estimated. In the simulation, the wet phase should evaluate not only the folding stabilities of RNA and peptide individually but also the overall stability of the RNA–peptide complex—complexes with higher stability survive degradation with higher probability and are preferentially utilized in subsequent cycles, thereby achieving simultaneous convergence of both sequence distributions.

Possibilities for experimental verification. Design experiments simulating wet-dry cycling, using high-throughput sequencing to directly detect sequence convergence and structural emergence in random RNA libraries after repeated wet-dry cycling. Reconstruct primitive cell systems in vitro to verify the osmotically driven material accumulation predicted by the membrane permeability hypothesis and population-level parallel selection. Further design experiments to test whether complexes formed by specific ribosomal proteins and high-frequency stable RNA fragments possess independent catalytic capacity for peptide bond formation or other key chemical reactions.

Implications for the search for extraterrestrial life. This theory predicts that the conditions for the origin of life are not stringent—as long as periodic environmental fluctuations (such as wet-dry cycles) and lipid molecules capable of forming selectively permeable membranes are present, life could spontaneously emerge on any Earth-like planet. This is a deterministic physicochemical process whose timescale can, in principle, be estimated, rather than a fortuitous miracle.

The emergence of complex life and intelligent life, however, requires stricter conditions: the simultaneous presence of land and ocean. Terrestrial hot springs provide the initial crucible for the origin of life; mild oceanic environments allow cells to escape the harsh wet-dry cycles and generate multiple selectable steady-state solutions under more stable conditions; and changing environmental conditions (tides, seasons, geological activity) continuously induce living systems to migrate and explore among multiple steady-state solutions. This combination of "multi-stability + environmental perturbation" is the key physical prerequisite for the eventual emergence of multicellular organisms and intelligent life. Hence, in the search for extraterrestrial life, simply detecting basic biomolecules (such as amino acids, nucleic acids) is insufficient—these molecules may be widely distributed throughout the universe. What is truly rare are planets that simultaneously possess terrestrial hot springs, liquid oceans, and periodic environmental changes (such as tides)—the combination of these conditions is the decisive astrophysical factor for the emergence of complex and intelligent life.

## 8.9 Chapter Summary and Outlook

Stability-distillation, as a unified cross-scale principle, persistently drives complex systems from randomness toward order—from individual molecules to cellular populations, from the origin of life to neural systems and artificial intelligence. The theory presented in this paper reveals the recurrent manifestation of this principle at multiple scales—the functional differentiation of index repositories and computers, the emergence and transitions of multi-stable systems, and the universal convergence mechanism of "attempt–evaluate–retain"—constituting a hierarchically nested self-organization framework. In the process of abiogenesis, as reactions proceed layer by layer, information content progressively increases, and essential information is stored in an index repository (DNA), enabling the system to more efficiently retrieve the information-processing procedure—this is fundamentally analogous to multi-level sequence representations.

This theory is not in competition with existing major hypotheses; rather, it integrates and refines them—providing a unified physicochemical foundation that preserves and harmonizes the rational kernels each has captured within a logically self-consistent framework. The testable predictions of this paper span multiple levels, from RNA and proteins to genomes and evolutionary biology. The abundance of non-coding RNAs and viral sequences in modern genomics has largely confirmed the core inference of "DNA as a mapping of environmental RNA"; the severe incompleteness of ribosomal protein genes in the LUCA genome provides fossil evidence for "gradual gene capture"; the independent catalytic functions of ribosomal proteins offer experimental support for "the ribosome originating from free catalytic modules."

The origin of life is a deterministic physicochemical process whose timescale can, in principle, be calculated, rather than a fortuitous miracle. But the emergence of complex life and intelligent life requires stricter astrophysical conditions—the simultaneous presence of land, ocean, and periodic environmental changes. This perspective not only reshapes our understanding of the origin of life but also provides clear guidance for the search for extraterrestrial life, and brings the chemical principles of life's origin under the same unified organizational principle as the cognitive architecture of neural systems and the cultural evolution of social systems.

## Acknowledgements

During the preparation of this work, the author (Cheng Bi) used DeepSeek to polish the language and improve readability. After using this tool, the author reviewed and edited the content as needed and takes full responsibility for the content of the publication. For their steadfast companionship during the writing of this manuscript, I am grateful to my cats—Xiao Ju, Orange, and Mie Mie.